\documentclass[final,number,sort&compress]{elsarticle}

\usepackage{amsmath,amsfonts,amssymb,amsbsy,amscd}
\usepackage{ifthen}
\usepackage{graphicx}
\usepackage[dvips]{color}
\usepackage[dvips,colorlinks]{hyperref}

\graphicspath{{figs/}} 
                            \newif\ifdraft \newif\ifpaper
 \draftfalse\paperfalse     % final version, hyperlinks

\ifpaper % prepare for B&W paper printing:
       
       \newcommand{\wwwcb}[1]{{\tt ChaosBook.org#1}}
       \newcommand{\arXiv}[1]{ {\tt arXiv:#1}}
\else % prepare hyperlinked pdf
        \newcommand{\wwwcb}[1]{   
                  {\tt \href{http://ChaosBook.org#1}
              {ChaosBook.org#1}}}
       
       \newcommand{\arXiv}[1]{
              {\tt \href{http://arXiv.org/abs/#1}{\goodbreak arXiv:#1}}}
\fi

\newcommand{\beq}{\begin{equation}}
\newcommand{\continue}{\nonumber \\ }
\newcommand{\nnu}{\nonumber}
\newcommand{\eeq}{\end{equation}}
\newcommand{\ee}[1]{\label{#1} \end{equation}}
\newcommand{\bea}{\begin{eqnarray}}

\newcommand{\eea}{\end{eqnarray}}
\newcommand{\barr}{\begin{array}}
\newcommand{\earr}{\end{array}}

\newcommand{\rf}     [1] {~\cite{#1}}
\newcommand{\refref} [1] {ref.~\cite{#1}}

\newcommand{\refrefs}[1] {refs.~\cite{#1}}

\newcommand{\refeq}  [1] {(\ref{#1})}

\newcommand{\reffig} [1] {figure~\ref{#1}}

\newcommand{\refFig} [1] {Figure~\ref{#1}}

\newcommand{\refsect}[1] {sect.~\ref{#1}}
\newcommand{\refsects}[2] {sects.~\ref{#1} and \ref{#2}}
\newcommand{\refSect}[1] {Sect.~\ref{#1}}

\newcommand\Poincare{Poincar\'e }
\newcommand{\statesp}{state space}
\newcommand{\Statesp}{State space}
\newcommand{\stabmat}{stability matrix}   
   
\newcommand{\jacobianM}{Jacobian matrix}  

\newcommand{\stretchf}{`stretch \&\ fold'}

\newcommand{\turn}{turning point}    % 
\newcommand{\Turn}{Turning point}    % 
\newcommand{\nws}{non--wandering set}
\newcommand{\po}{periodic orbit}

\newcommand{\rpo}{relative periodic orbit}
\newcommand{\Rpo}{Relative periodic orbit}
\newcommand{\eqv}{equilibrium}

\newcommand{\eqva}{equilibria}

\newcommand{\reqv}{relative equilibrium}

\newcommand{\reqva}{relative equilibria}
\newcommand{\Reqva}{Relative equilibria}
\newcommand{\reducedsp}{reduced state space}
\newcommand{\Reducedsp}{Reduced state space}
\newcommand{\fixedsp}{fixed-point subspace}

\newcommand{\slice}{slice}
\newcommand{\Slice}{Slice}
\newcommand{\mslices}{method of slices}

\newcommand{\mframes}{{method of moving frames}}

\newcommand{\cLe}{complex Lorenz equations}
\newcommand{\cLf}{complex Lorenz flow}
\newcommand{\CLe}{Complex Lorenz equations}

\newcommand{\KS}{Kuramoto-Sivashinsky}
\newcommand{\KSe}{Kuramoto-Sivashinsky equation}
\newcommand{\pCf}{plane Couette flow}

\newcommand{\etc}{{etc.}}       
\newcommand{\etal}{{\em et al.}}
\newcommand{\ie}{{i.e.}}

\newcommand{\reals}{\mathbb{R}}
\newcommand{\Rls}[1]{\ensuremath{\mathbb{R}^{#1}}}
\newcommand{\pde}{\partial}
\newcommand {\id}{{\ \hbox{{\rm 1}\kern-.6em\hbox{\rm 1}}}}
\newcommand{\pS}{\ensuremath{{\cal M}}} 
\newcommand{\ssp}{\ensuremath{x}}       
\newcommand\xInit{{x_0}}        
\newcommand\flow[2]{{f^{#1}(#2)}}
\newcommand{\vel}{\ensuremath{v}}  
\newcommand{\Mvar}{\ensuremath{A}} 
\newcommand{\Lyap}{\ensuremath{\lambda}}           
\newcommand{\eigExp}[1][]{
     \ifthenelse{\equal{#1}{}}{\ensuremath{\lambda}}{\ensuremath{\lambda^{(#1)}}}}
\newcommand{\eigRe}[1][]{
     \ifthenelse{\equal{#1}{}}{\ensuremath{\mu}}{\ensuremath{\mu^{(#1)}}}}
\newcommand{\eigIm}[1][]{
     \ifthenelse{\equal{#1}{}}{\ensuremath{\omega}}{\ensuremath{\omega^{(#1)}}}}
\newcommand{\PoincS}{\ensuremath{\cal P}}     
\newcommand{\PoincM}{\ensuremath{P}}       
\newcommand{\tr}{\mbox{\rm tr}\,}

\newcommand\stagn{q}      
\newcommand{\cycle}[1]{\ensuremath{\overline{#1}}}
\newcommand\period[1]{{\ensuremath{T_{#1}}}}         
\newcommand{\EQV}[1]{\ensuremath{EQ_{#1}}} 
\newcommand{\REQV}[2]{\ensuremath{TW_{#1#2}}}

\newcommand{\rLor}{\rho}  
\newcommand{\RerCLor}{\rho_1}
\newcommand{\ImrCLor}{\rho_2}

\newcommand{\gSpace}{\ensuremath{{\bf \theta}}}
\newcommand{\stab}[1]{\ensuremath{G_{#1}}}
\newcommand{\velRel}{\ensuremath{c}}    
\newcommand{\pVeloc}{v}         
\newcommand{\Fix}[1]{\ensuremath{\mathrm{Fix}\left(#1\right)}}
\newcommand{\pSRed}{\ensuremath{\overline{\cal M}}}
\newcommand{\sspRed}{\ensuremath{y}}    
\newcommand{\velRed}{\ensuremath{u}}    
\newcommand{\slicep}{{\ensuremath{y'}}} 
\newcommand{\sliceTan}[1]{\ensuremath{t'_{#1}}}
\newcommand{\groupTan}{\ensuremath{t}}   
\newcommand{\Group}{\ensuremath{G}}      
\newcommand{\Lg}{\ensuremath{\mathbf{T}}}
\newcommand{\LieEl}{\ensuremath{g}} 
\newcommand{\Un}[1]{\ensuremath{\textrm{U}(#1)}}
\newcommand{\On}[1]{\ensuremath{\textrm{O}(#1)}}
\newcommand{\SOn}[1]{\ensuremath{\textrm{SO}(#1)}}

\newcommand{\vf}{v}	

\newcommand{\ode}{ODE}

\newcommand{\REQB}[1]{\ensuremath{\mathrm{Q}_{#1}}} 
\renewcommand{\REQV}[2]{\ensuremath{\mathrm{Q}_{#1#2}}} 
\renewcommand{\EQV}[1]{\ensuremath{\mathrm{E}_{#1}}} 

\renewcommand\Poincare{Poincar\'e}

\renewcommand{\sspRed}{\ensuremath{\overline{x}}}    
\renewcommand{\slicep}{{\ensuremath{\overline{x}'}}} 
\renewcommand{\sliceTan}[1]{\ensuremath{t'_{#1}}}    
\renewcommand{\groupTan}{\ensuremath{t}}

\newcommand{\slicepComp}[2]{{\ensuremath{\overline{#1}'_{#2}}}} 
\newcommand{\Subgroup}{H}

\newcommand{\sset}{singular set}

\newcommand{\cont}{\,, \\ }

\begin{document}
\journal{Physica D}
\begin{frontmatter}

			\title{
Continuous symmetry reduction and return maps for high-dimensional flows
			}

\author{Evangelos Siminos}
\ead{siminos@gatech.edu}
\author{Predrag Cvitanovi\'c}
\address{Center for Nonlinear Science,
        School of Physics, Georgia Institute of Technology,
        Atlanta, GA 30332-0430}

        \begin{abstract}
We present two continuous symmetry reduction methods for
reducing high-dimensional dissipative flows to local return
maps. In the Hilbert polynomial basis approach, the equi\-vari\-ant
dynamics is rewritten in terms of in\-vari\-ant coordinates. In the
method of moving frames (or method of slices) the state space
is sliced locally in such a way that each group orbit of
symmetry-equivalent points is represented by a single point. In
either approach, numerical computations can be performed in the
original state-space representation, and the solutions are then
projected onto the symmetry-reduced state space. The two methods
are illustrated by reduction of the complex Lorenz system, a
5-dimensional dissipative flow with rotational symmetry. While
the Hilbert polynomial basis approach appears unfeasible for
high-dimensional flows, symmetry reduction by the method of
moving frames offers hope.
        \end{abstract}

\end{frontmatter}

\section{\label{s:intro} Introduction}

In his seminal paper, E. Lorenz\rf{lorenz} reduced the
continuous time and discrete spatial symmetries of the
3-dimensional Lorenz equations, resulting in a 1-dimensional
return map that yields deep insights\rf{tucker1-2} into the
nature of chaos in this flow. For strongly contracting,
low-dimensional flows, Gilmore, Lefranc and
Letellier\rf{gilmore2003,GL-Gil07b} systematized construction
of such discrete time return maps, through use of topological
templates, Poincar\'e sections (to reduce the continuous time
invariance) and in\-vari\-ant polynomial bases (to reduce the
spatial symmetries). They showed that in presence of spatial
symmetries one has to  `quotient' the symmetry and replace
the dynamics by a physically equivalent reduced,
desymmetrized flow, in which each family of symmetry-related
states is replaced by a single representative. This approach
leads to symbolic dynamics and labeling of all \po s up to a
given topological period. Periodic orbit theory can then
yield accurate estimates of long-time dynamical averages,
such as Lyapunov exponents and escape rates\rf{DasBuch}.

In a series of papers Cvitanovi\'{c}, Putkaradze,
Christiansen and Lan%
\rf{Christiansen97,chfield,LanThesis,CvitLanCrete02,lanVar1,lanCvit07}
showed that effectively low-dimensional return maps can be
constructed for high-dimensional (formally infinite
dimensional)  flows described by dissipative partial
differential equations (PDEs) such as the \KSe\ (KS). Such
flows have state-space topology vastly more complicated than
the Lorenz flow, and collections of local Poincar\'e sections
together with maps from a section to a section are required
to capture all of the important asymptotic dynamics. These KS
studies were facilitated by a restriction to the
flow-in\-vari\-ant subspace of odd solutions, but at a price:
elimination of the translational symmetry of the KS system
and with it physically important phenomena, such as traveling
waves. Traveling (or relative) unstable coherent solutions
are ubiquitous and play a key role in organization of
turbulent hydrodynamic flows, as pointed already in
1982 by Rand\rf{Rand82}, and confirmed both by
simulations and experimentation%
\rf{KawKida01,FE03,WK04,Visw07b,GHCW07,science04}.
For KS\rf{SCD07,SiminosThesis}, and even for a relatively
low-dimensional flow such as the
\cLe\rf{GibMcCLE82,FowlerCLE82} used as an example here, with
the simplest possible continuous (rotational) spatial
symmetry, the symmetry-induced drifts obscure the underlying
hyperbolic dynamics.

The question that we address here is how one can construct
suitable return maps for arbitrarily high-dimensional but
strongly dissipative flows in presence of continuous
symmetries. Our exposition is based in part on
\refrefs{SiminosThesis,DasBuch,Wilczak09}. The reader is
referred to Golubitsky and
Stewart\rf{golubitsky2002sp}, Hoyle\rf{hoyll06},
Olver\rf{OlverInv}, Bredon\rf{Bredon72}, and
Krupa\rf{Krupa90} for more depth and rigor than would be wise
to wade into here.

In \refsect{s:symDyn} we review the basic notions of symmetry
in dynamics. \refSect{s:introCLE} introduces the \SOn{2}\
equi\-vari\-ant \cLe\ (CLE), a 5-dimen\-sion\-al set of ODEs that we
use throughout the paper to illustrate the strengths and
drawbacks of different symmetry reduction methods. In
\refsect{s:symSol} we describe important classes of solutions
and their symmetries: \eqva, \reqva, periodic and \rpo s, and use
them to motivate the need for symmetry reduction.

In \refsect{s:symmRed} we describe the problem of
\emph{symmetry reduction}. The action of a symmetry group
endows the \statesp\ with the structure of a union of group
orbits, each group orbit an equivalence class. The goal of
{symmetry reduction} is replace each group orbit by a unique
point a lower-dimensional {\em \reducedsp}.
In \refsect{s:Hilbert} we briefly review the standard approach
to spatial symmetry reduction, projection to a Hilbert basis,
and explain why we find it impracticable.
In \refsect{sec:mf} we review the {\mframes}, a direct and
efficient method  for computing symmetry-in\-vari\-ant bases that
goes back to Cartan, and in \refsect{sec:CLeMovFr} we apply
the method to the \cLe. The method maps all solutions to a
slice, a submanifold  of state space that plays a role for
group orbits akin to the role Poincar\'e sections play in
reducing continuous time invariance. In contrast to the
Hilbert basis approach, slices are local, with a generic
trajectory within a slice bound to encounter singularities,
and more than one slice might be needed to capture the flow
globally. In \refsect{s:cleCoordSlice} we show that a single
local slice can suffice for the purpose of reducing the \cLe\
flow to a return map. In \refsect{sec:MovFrameODE} we recast
the {\mframes} into the equivalent, differential \mslices,
with time integration restricted to a slice fixed by a given
\statesp\ point.

\section{\label{s:symDyn} Symmetries of dynamical systems}

Here we are interested in the role continuous symmetries
play in dynamics.
The methods we develop
are in principle applicable to
to translational and rotational symmetries of ODEs and
PDEs, described by compact or noncompact Lie groups.
We have in mind applications to PDEs such as \KS\
and \pCf\ which exhibit translational symmetries in either infinite
or periodic domains. In the former case the group of symmetries
is Euclidean and noncompact, and in the latter case
orthogonal and compact.
In numerical computations
the periodic setting is usually considered and, though Fourier analysis,
a translation is represented by the action of the 1-parameter Lie \SOn{2} group
on its linearly irreducible subspaces, the Fourier modes.
Through truncation (for example spectral discretization),
PDEs are transformed to high- but finite-dimensional systems of \ode s.
The key concepts will thus
be illustrated by a specific ODE example, the \SOn{2} group acting
on a five-dimensional state space, linearly decomposable into
a direct sum of irreducible subspaces of \SOn{2}.

Consider a system of \ode s of the form
\beq
	\dot{\ssp} = \vf(\ssp)
	\label{eq:difeq}
\eeq
with $\vf$ a smooth vector field and $\ssp\in\pS\subset\Rls{d}$.

A linear action $\LieEl$ is a symmetry of
\refeq{eq:difeq} if
\beq
	\vf(\LieEl \ssp) =\LieEl \, \vf(\ssp)
	\label{eq:equiv}
\eeq
for all $\ssp\in\Rls{d}$. One  says that $\vf$ \emph{commutes}
with $\LieEl$ or that $\vf$ is $\LieEl$-\emph{equi\-vari\-ant}.
When $\vf$ commutes with the set of group elements
$\LieEl\in\Group$, the vector field $\vf$ is said to be
$\Group$-equi\-vari\-ant. The group $\Group$ is said to be a {\em
symmetry} of dynamics if for every solution $\ssp(\tau)=
\flow{\tau}{\ssp}$, $\LieEl \, \ssp(\tau)$ is also a solution. The finite
time flow $\flow{\tau}{\LieEl \ssp}$ through $\LieEl \ssp$ then
satisfies the equivariance condition
\beq
\flow{\tau}{\LieEl \ssp}=\LieEl\flow{\tau}{\ssp}
\,.
\ee{eq:equivFinite}
In physics literature the term $in\-vari\-ant$ is most commonly
used; for example, in Hamiltonian systems a symmetry is
manifested as invariance of the Hamiltonian under the
symmetry group action.

An element of a compact Lie group
continuously connected to identity can be written as
\beq
\LieEl(\gSpace)=e^{\gSpace \cdot \Lg }
	\,,\qquad
\gSpace \cdot \Lg  = \sum \gSpace_a \Lg_a,\; a=1,2, \cdots, N
\,,
\ee{FiniteRot}
where
$\gSpace \cdot \Lg$
is a {\em Lie algebra} element,  and $\gSpace_a$ are the parameters
of the transformation. Repeated indices are summed throughout this
chapter, and the dot product refers to a sum over
Lie algebra generators. The Euclidian product of two vectors
$x,y$ is indicated by $x$-transpose times $y$, \ie,
$x^T y = \sum_i^d x_i y_i$.
Finite transformations $ \exp(\gSpace \cdot {\Lg}) $ are
generated by sequences of infinitesimal steps of form
\beq
\LieEl(\delta\gSpace) \simeq 1 + \delta \gSpace \cdot \Lg
    \,,\quad
\delta\gSpace \in \reals^N
    \,,\quad
|\delta \gSpace| \ll 1
    \, ,
\ee{intsmLieTransf}
where $\Lg_a$, the {\em generators} of infinitesimal
transformations, are a set of $N$ linearly independent
$[d\!\times\!d]$ anti-hermitian matrices, $(\Lg_a)^\dagger =
- \Lg_a$, acting linearly on the $d$-dim\-ens\-ion\-al \statesp\
$\pS$.
For $\Group \subset \On{n}$ the generators can
always be brought to real, antisymmetric form
$\Lg^T=-\Lg$.
The flow induced by the action of the group
on the \statesp\ point $\ssp$ 
is given by the set of $N$ tangent fields
\beq
\groupTan_a(\ssp)_{i}= (\Lg_a){}_{ij} \ssp_j
\,.
\ee{GroupTangField}
These tangent fields are always normal to the
`radial' vector $\ssp$,
\beq
\ssp^T \groupTan_a(\ssp)= 0
\,.
\ee{TangFieldNormal}

For an infinitesimal transformation \refeq{intsmLieTransf}
the $\Group$-equivariance condition \refeq{eq:equiv}
becomes
\[
\vel(\ssp)
      \simeq
  (1-\gSpace \cdot \Lg) \, \vel(\ssp+\gSpace \cdot \Lg \, \ssp)
       = \vel(\ssp)- \gSpace \cdot \Lg \, \vel(\ssp)
             + \frac{d\vel}{d\ssp} \,\gSpace \cdot \Lg \, \ssp
\,.
\]
Thus the
infinitesimal, Lie algebra $\Group$-equivariance condition is
\beq
  \groupTan_a(\vel)  - \Mvar(\ssp) \, \groupTan_a(\ssp) =0
  \,,
\ee{inftmInv}
where $\Mvar = {\pde \vel}/{\pde \ssp}$ is the \stabmat.
The left-hand side,
\beq
{\cal L}_{\groupTan_a} \vel =
\left.\left(
  \Lg_a - \frac{\partial}{\partial y}(\Lg_a \ssp)
 \right) \vel(y)\right|_{y=\ssp}
 \,,
\ee{LieDeriv}
is known as
the {\em Lie derivative} of the dynamical flow
field $\vel$ along the direction of the infinitesimal
group-rotation induced flow $\groupTan_a(\ssp)= \Lg_a \ssp$.
The equivariance condition \refeq{inftmInv} states that the two
flows, one induced by the dynamical vector field $\vel$, and
the other by the group tangent field $\groupTan$, commute if
their Lie derivatives (or the Lie brackets or Poisson
brackets) vanish.

Any representation of a compact Lie group $\Group$ is fully
reducible, and in\-vari\-ant tensors constructed by contractions
of $\Lg_a$ are useful for identifying irreducible
representations. The simplest such in\-vari\-ant is
bilinear,
\beq
\Lg^T \cdot \Lg = \sum_\alpha C_2^{(\alpha)} \, \id^{(\alpha)}
\,,
\ee{QuadCasimir}
where $C_2^{(\alpha)}$ is the quadratic Casimir for
irreducible representation labeled $\alpha$, and
$\id^{(\alpha)}$ is the identity on the $\alpha$-irreducible
subspace, 0 elsewhere. The dot product of two tangent fields
is thus a sum weighted by Casimirs,
\beq
\groupTan(\ssp)^T  \cdot \groupTan(\ssp')
   = \sum_\alpha C_2^{(\alpha)} \ssp_i\, \delta_{ij}^{(\alpha)} \ssp'_j
\,.
\ee{dotProd}
If $\ssp$ is not invariant (fixed under group actions), 
$\groupTan(\ssp)^T  \cdot \groupTan(\ssp)$ is strictly positive. 
$\groupTan(\ssp)^T  \cdot \groupTan(\ssp')$, however, can take either sign, or even
vanish.
    
\subsection{\label{s:introCLE} An example: \CLe}

Consider a complex generalization of Lorenz equations,
\bea
 \dot{x} &=& -\sigma x+ \sigma y \,,\qquad
 \dot{y} \,=\, (\rLor-z)x-a y \continue
 \dot{z} &=& (x y^*+x^*y)/2 -b z\,,
 \label{eq:CLe}
\eea
where $x,y$ are complex variables, $z$ is real, while the
parameters $\sigma,\,b$ are real and $\rLor=\RerCLor+i
\ImrCLor$, $a=1-i e$ are complex. Recast in real variables,
$x=x_1+ i x_2\,,\,y=y_1+ i y_2$ this is a set of five coupled ODEs
\bea
	\dot{x}_1 &=& -\sigma x_1 + \sigma y_1
            \,,\quad
	\dot{x}_2 \,=\, -\sigma x_2 + \sigma y_2\continue
	\dot{y}_1 &=& (\RerCLor-z) x_1 - \ImrCLor x_2 -y_1-e y_2 \continue
	\dot{y}_2 &=& \ImrCLor x_1 + (\RerCLor-z) x_2 + e y_1- y_2\continue
	\dot{z} \; &=& -b z + x_1 y_1 + x_2 y_2
    \,.
\label{eq:CLeR}
\eea
In all numerical examples that follow, the parameters will be
set to $\RerCLor=28,\, \ImrCLor=0,\, b=8/3,\, \sigma=10,\, e=
1/10$, unless explicitly stated otherwise.
%
%%%%%%%%%%%%%%%%%%%%%%%%%%%%%%%%%%%%%%%%%%%%%%%%%%%%%%%%%%%%
\begin{figure}[ht]
\begin{center}
  \includegraphics[width=0.40\textwidth, clip=true]{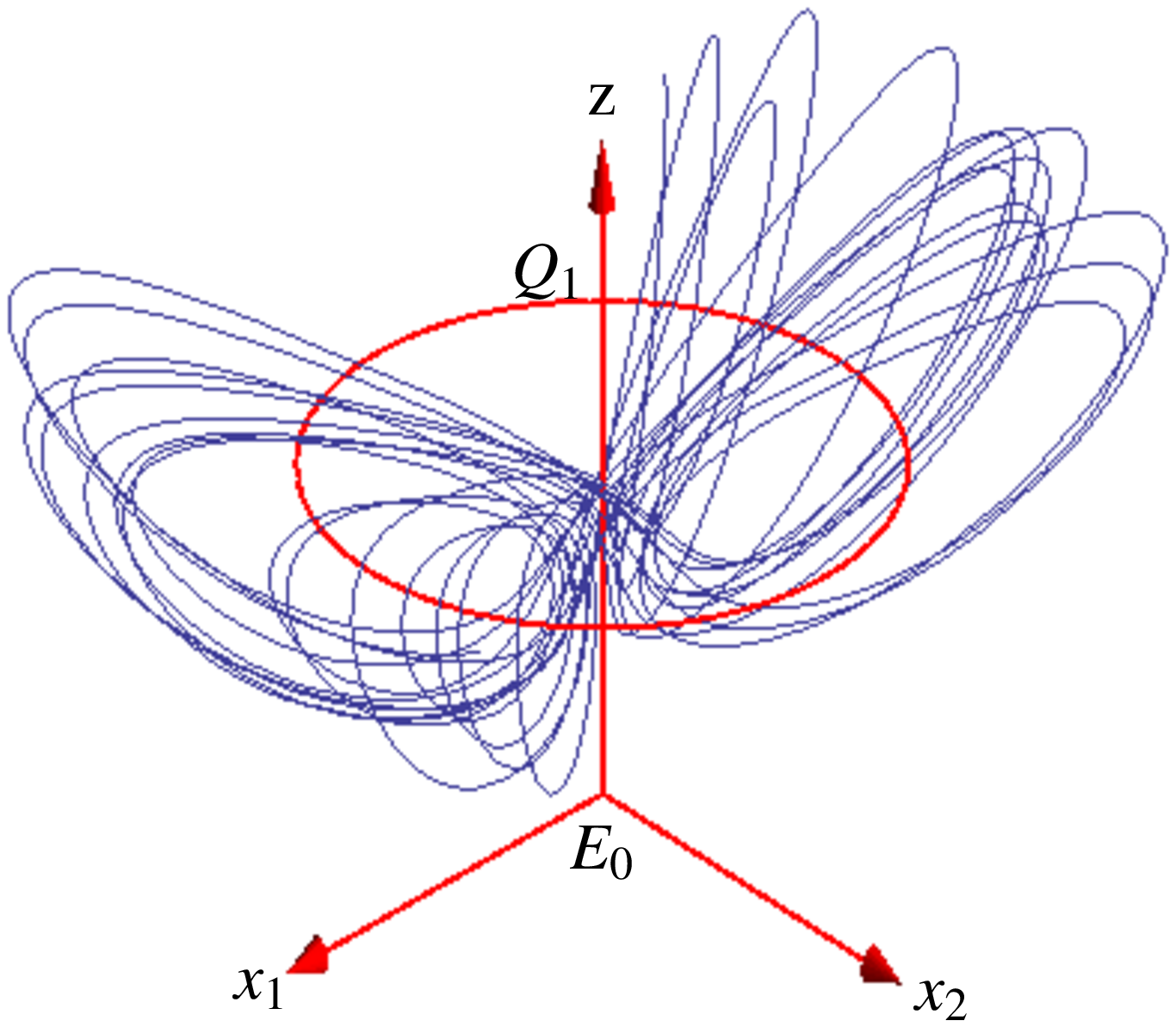}
  \includegraphics[width=0.40\textwidth, clip=true]{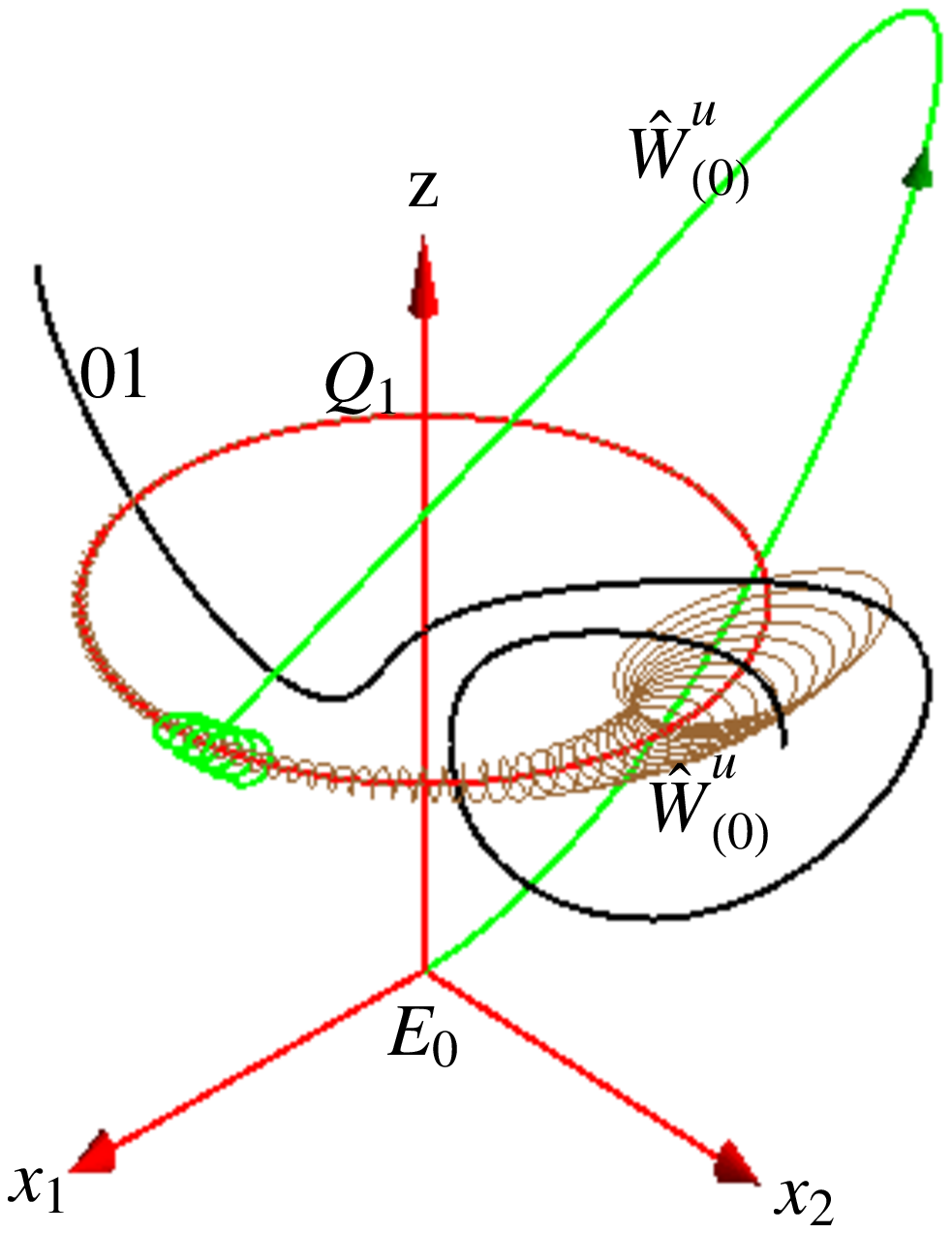}
\end{center}
\caption{
\Statesp\ portrait of \cLf. Plotted are a generic chaotic trajectory (blue),
the \EQV{0} \eqv,
a representative of its unstable manifold (green),
the \REQV{}{1} \reqv\ (red), its unstable manifold (brown), and
one repeat of the \cycle{01} \rpo\ (black) (color online).
}
\label{fig:CLE}
\end{figure}
%%%%%%%%%%%%%%%%%%%%%%%%%%%%%%%%%%%%%%%%%%%%%%%%%%%%%%%%%%%%
%
Why worry about continuous symmetries? The visualization
in \reffig{fig:CLE} of typical long-time dynamics of \cLf\ suffices
to illustrate the effect a continuous symmetry has on
dynamics. A generic trajectory slowly drifts along the
direction of continuous symmetry while tracing a
Lorenz-butterfly like attractor. It is a mess.

The \cLe\ are a dynamical system with a continuous
(but no discrete) symmetry, equi\-vari\-ant under the one-parameter
rotation group $\Un{1}\cong\SOn{2}$ acting by
\beq\label{eq:SO2cle}
	(x,\,y,\,z)\mapsto (
    e^{i\theta}x,\,e^{i\theta}y,\,z)\,,\ \theta\in[0,2\pi]
\,.
\eeq
Alternatively, substituting the Lie algebra generator
\beq
 \Lg \,=\,   \left(\barr{ccccc}
    0  & -1 & 0  &  0 & 0  \\
    1  &  0 & 0  &  0 & 0 \\
    0  &  0 & 0  & -1 & 0  \\
    0  &  0 & 1  &  0 & 0 \\
    0  &  0 & 0  &  0 & 0
    \earr\right)
\ee{CLfLieGen}
acting on a 5-dim\-ens\-ion\-al space \refeq{eq:CLeR} into
\refeq{FiniteRot} yields the  $\Rls{5}$ representation of a
finite angle action \refeq{eq:SO2cle} of $\SOn{2}$
\beq
\LieEl(\gSpace) \,=\,  \left(\barr{ccccc}
  \cos \gSpace  & -\sin \gSpace  & 0 & 0 & 0 \\
  \sin \gSpace  &  \cos \gSpace  & 0 & 0 & 0 \\
 0 & 0 &  \cos \gSpace & -\sin \gSpace   & 0 \\
 0 & 0 &  \sin \gSpace &  \cos \gSpace   & 0 \\
 0 & 0 & 0             & 0               & 1
    \earr\right)
\,.
\ee{CLfRots}
We see that the linear action of \SOn{2}\
on the \statesp\ of the \cLe\
decomposes into the $m\!=\!0$ \Group-in\-vari\-ant
subspace ($z$-axis) and  the $m=1$ subspace of multiplicity 2.

The generator $\Lg$ is
anti-symmetric,
$\Lg^T = - \Lg$, 
and the group is compact, its
elements parametrized by $\gSpace \mbox{ mod } 2\pi$. Locally, at
$\ssp \in \pS$, the infinitesimal action of the group is
given by the group tangent field $\groupTan(\ssp) = \Lg \ssp
= (-x_2,x_1,-y_2,y_1,0)$. In other words, the flow induced by
the group action is normal to the radial direction in the
$(x_1,x_2)$ and $(y_1,y_2)$ planes, while the $z$-axis is left
in\-vari\-ant.

The equivariance of the \cLf\ under $\SOn{2}$ rotations
\refeq{CLfRots} can be verified
by substituting the Lie algebra generator
\refeq{CLfLieGen} and the \stabmat\ for \cLf\ \refeq{eq:CLeR},
  \beq
\Mvar =
  \left(\barr{ccccc}
    -\sigma    	& 0 		& \sigma & 0    &  0 \\
	0 	& -\sigma       & 0      & \sigma   &  0 \\
	\RerCLor-z  &     -\ImrCLor      & -1     & -e & -x_1 \\
	\ImrCLor     & \RerCLor-z       	& e  	& -1       & -x_2 \\
	y_1     & y_2           & x_1    & x_2      & -b
    \earr\right)
\,,
  \ee{CLeStabMat}
into the equivariance condition \refeq{inftmInv}.
For the parameter values \refeq{eq:CLeR} the flow
is strongly volume contracting,
\beq
\pde_i \pVeloc_i
 = \tr \Mvar
 = \sum_{i=1}^{5} \Lyap_i(\ssp,t)
= -b -2(\sigma + 1)
= -24 - 2/3
    \,.
\ee{trA-ZM}

The \cLe\ \refeq{eq:CLe} were introduced by
Gibbon and McGuinness\rf{GibMcCLE82,FowlerCLE82} as a
low-dim\-ens\-ion\-al model of baroclinic instability in the
atmosphere. Zeghlache and Mandel\rf{ZeMa85} and
Ning and Haken\rf{NingHakenCLE90} have shown that
equations isomorphic to the \cLe, with $e+\ImrCLor=0$,
also appear as a truncation of Maxwell-Bloch equations
describing a single mode, detuned, ring laser. The choice
$e+\ImrCLor=0$ is degenerate
(see \refeq{eq:omegaCLE}) in the sense that
it leads to non-generic bifurcations. We follow Bakasov and
Abraham\rf{BakasovAbraham93} who set $\ImrCLor=0$ and $e \neq
0$ to describe detuned ring lasers.

Here, however, we are not interested in the physical
applications of these equations; rather, we study them as a
simple example of a dynamical system with continuous (but no
discrete) symmetries, with a view of testing methods of
reducing the dynamics to a lower-dimensional \reducedsp.
We investigate
various ways of quotienting its \SOn{2} symmetry, and
reducing the dynamics to a 4-dim\-ens\-ion\-al \reducedsp. As
we shall show, the dynamics has a nice {\stretchf}
action, but that is totally masked by the continuous symmetry
drifts. We shall not rest until we attain the simplicity of
\reffig{fig:CLEmf}, and the bliss of 1-dim\-ens\-ion\-al
return map of \reffig{fig:CLEip}.

\section{\label{s:symSol} Symmetries of solutions}

In order to explore the implications of equivariance on
solutions of dynamical equations,  we start by examining the
way a compact Lie group acts on a \statesp\ \pS. The
\emph{group orbit} or \emph{$\Group$-orbit} of the point
$\ssp \in \pS$ is the set
\beq
    \pS_\ssp = \{\LieEl\,\ssp \mid \LieEl \in {\Group}\}
\ee{GroupOrb}
of all \statesp\ points into which $\ssp$ is mapped under the
action of $\Group$.
The \emph{symmetry} $\stab{\ssp}$ (\emph{isotropy} or
\emph{stabilizer} group) of a \statesp\ point $\ssp$ is the
largest subgroup of $\Group$
\beq
\stab{\ssp} =\{\LieEl \in \Group: \LieEl \ssp = \ssp \}
\ee{def:isotr}
that leaves $\ssp$ fixed.
The \emph{symmetry} $\stab{X}$ of a set $\pS_X \in \pS$ is
the largest subgroup  of $\Group$ that leaves $\pS_X$
in\-vari\-ant as a set:
\[
	\stab{X}= \{\LieEl: \LieEl \, \pS_X = \pS_X\}
\,.
\]
If $\stab{p}$ is a symmetry, intrinsic properties of a
solution $\pS_p$ (such as \eqv\ or a cycle stability
eigenvalues, period, Floquet multipliers) evaluated anywhere
along its $\stab{p}$-orbit are the same. A symmetry thus
reduces the number of inequivalent solutions. So we also need
to describe the symmetry of a \emph{solution}, as opposed to
\refeq{eq:equivFinite}, the symmetry of the \emph{system}.

The \emph{\fixedsp} $\Fix{\Subgroup}$ of a subgroup
$\Subgroup\subset\Group$ is the subspace of $\pS$ containing
all fixed points of $\Subgroup$:
\[
	\Fix{\Subgroup}=
      \{\ssp\in\pS,\,\LieEl\in\Subgroup \,|\,
        \LieEl \ssp = \ssp \}
\,.
\]
The physical importance of \fixedsp s lies in the fact that
they are in\-vari\-ant under $\Group$-equi\-vari\-ant
dynamics\rf{golubitsky2002sp},
\[
 f^\tau\left(\Fix{\Subgroup}\right)\subseteq \Fix{\Subgroup}
\]
and thus \emph{flow in\-vari\-ant} for all times $\tau$.
Therefore if $\ssp(\tau)$ is a solution of an equi\-vari\-ant ODE,
then its symmetry $\stab{\ssp(\tau)}=\stab{\ssp(0)}$ is
preserved for all times.

%
%%%%%%%%%%%%%%%%%%%%%%%%%%%%%%%%%%%%%%%%%%%%%%%%%%%%%%%%%%%%%%%%
\begin{figure}[ht]
 (\textit{a})\includegraphics[width=0.40\textwidth,clip=true]{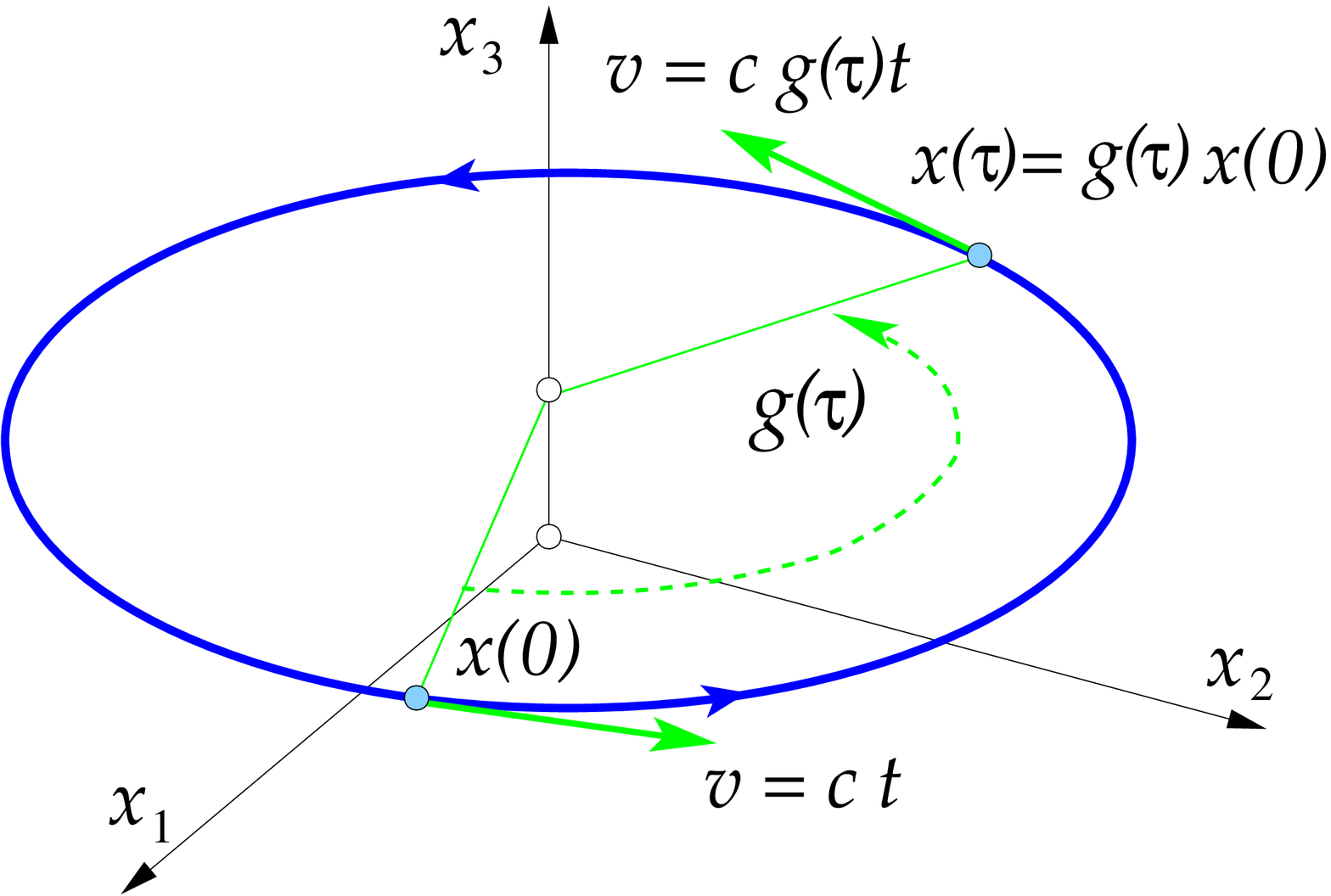}
~(\textit{b})\includegraphics[width=0.40\textwidth,clip=true]{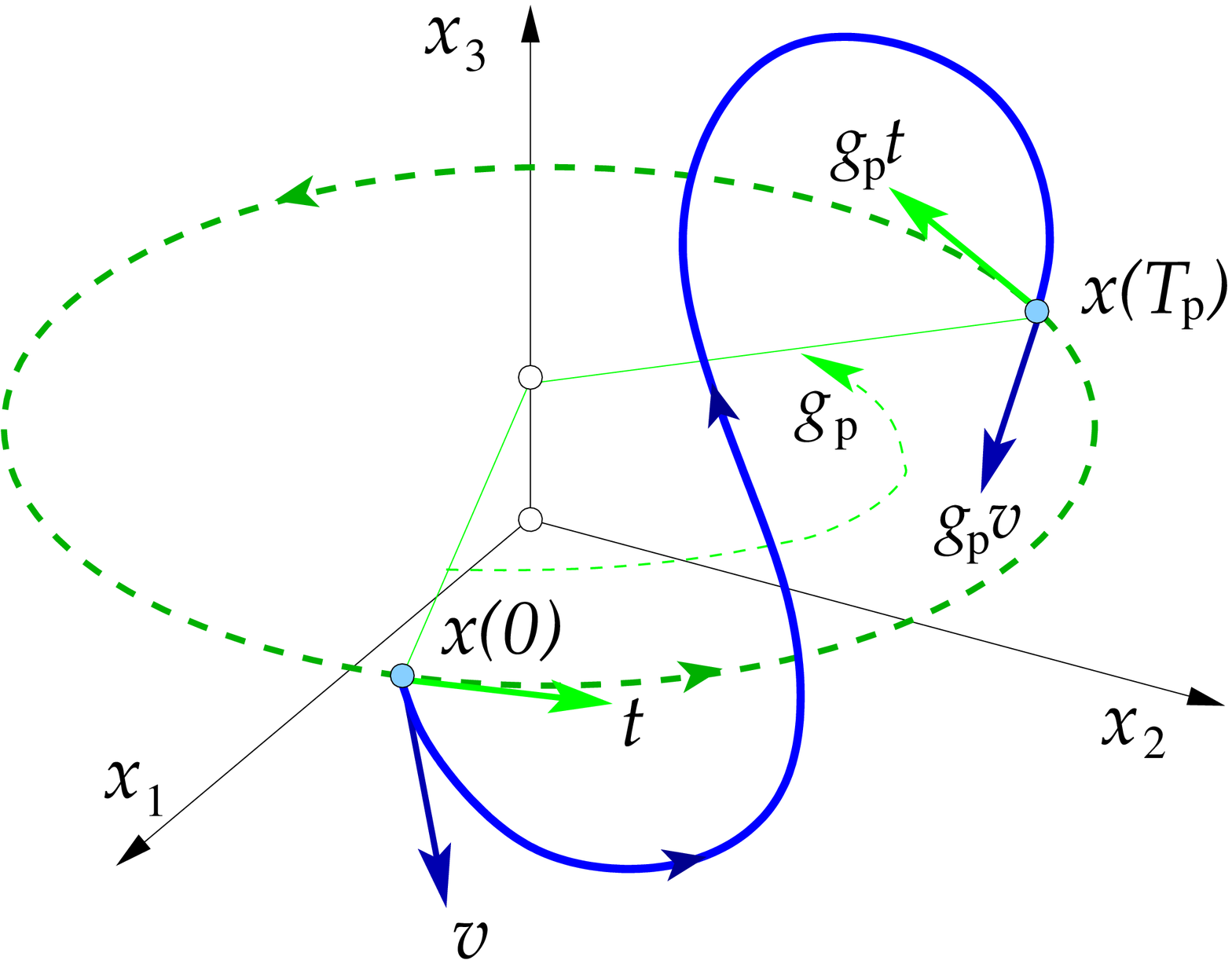}
\caption{
(a) A {\em \reqv\ orbit} starts at some point $\ssp(0)$,
with the dynamical flow field $\vel(\ssp) = \velRel \cdot
\groupTan(\ssp)$ pointing along the group tangent space. For
the $\SOn{2}$ symmetry depicted here, the flow traces out the
group orbit of $\ssp(0)$ in time $\period{}=2\pi/\velRel$.
An
{\em \eqv} lives either in the $\Fix{\Group}$ subspace
($x_3$ axis in this sketch), or on a group orbit as the one
depicted here, but with zero angular velocity $\velRel$. In
the latter case the circle (in general, $N$-torus) depicts a
continuous family of fixed \eqva, related only by the group
action.
(b) A {\em \rpo} starts out at $\ssp(0)$ with the dynamical $\vel$ and
group tangent $\groupTan$ flows pointing in different
directions, and returns to the group orbit of $\ssp(0)$ after
time $\period{p}$ at $\ssp(\period{p})=\LieEl_p \ssp (0)$, a
rotation of the initial point by $\LieEl_p$.
}
\label{f:rpo}
\end{figure}
%%%%%%%%%%%%%%%%%%%%%%%%%%%%%%%%%%%%%%%%%%%%%%%%%%%%%%%%%%%

In contrast to \emph{\eqv} solutions that satisfy
$f^\tau(\ssp)  =  \ssp$, \emph{\reqva} (or \emph{traveling
waves}) satisfy $f^\tau(\ssp) = \LieEl( \tau) \, \ssp$ for
any $\tau$, where the group has been reparameterized by
time, $\theta=\theta(\tau)$.
In a co-moving frame moving along the group orbit
with velocity $\vel(\ssp) = \velRel \cdot \groupTan(\ssp)$,
the \reqv\ appears as an \eqv. Here $\groupTan$ is the
group tangent field \refeq{GroupTangField}.

A {\em \rpo} is an orbit $\pS_p$ for which the initial point
exactly recurs
\beq
\ssp_p (0) = \LieEl_p \ssp_p (\period{p} )
    \,,\qquad
\ssp_p (\tau) \in \pS_p
    \,,
\label{RPOrelper1}
\eeq
at a fixed {\em relative period} $\period{p}$, but shifted by
a fixed group action ${\LieEl_p}$ which brings the endpoint
$\ssp_p (\period{p} ) $ back into the initial point $\ssp_p
(0) $, see \reffig{f:rpo}\,(b). The group action ${\LieEl_p}=
\LieEl_p(\gSpace)$ parameters $\gSpace_p =
(\gSpace_1,\gSpace_2,\cdots\gSpace_N)$ will be referred to as
phases, or shifts. For dynamical systems with only
continuous (no discrete) symmetries, the parameters
$\{t,\gSpace_1,\cdots,\gSpace_N\}$ are real numbers, the ratios
$\pi/\gSpace_j$ are almost never rational, and the likelihood
of closing into a {\po} is {zero}. Thus the trajectory
of a \rpo\ generically sweeps out the group orbit ergodically.

%
%%%%%%%%%%%%%%%%%%%%%%%%%%%%%%%%%%%%%%%%%%%%%%%%%%%%%%%%%%%%
\begin{figure}[ht] \label{f:MeanVelocityFrame}
(\textit{a})\includegraphics[width=0.40\textwidth, clip=true]
                    {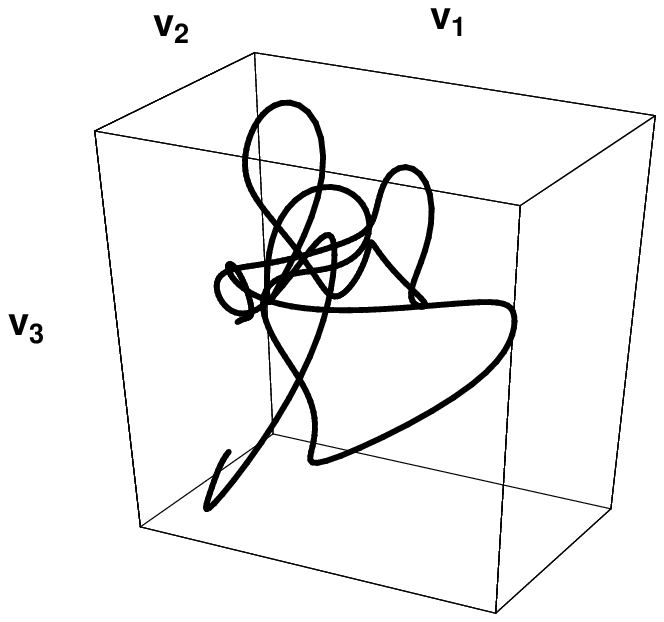}
~(\textit{b})\includegraphics[width=0.40\textwidth, clip=true]
                     {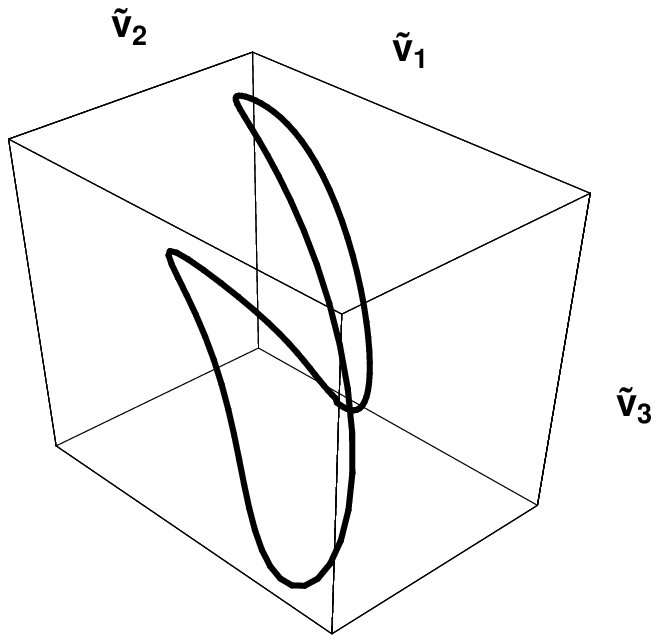}
\caption{
 A \rpo\ of the Kuramoto-Sivashinsky flow, traced for four periods
 $\period{p}$ and projected on
 (a) a stationary \statesp\ coordinate frame
 $\{v_1,v_2,v_3\}$;
 (b) a co-moving $\{\tilde{v}_1,\tilde{v}_2,\tilde{v}_3\}$
 coordinate frame, moving with the mean velocity
 $\velRel_p=\gSpace_p/\period{p}$.
(From \refref{SCD07}.)
}
\end{figure}
%%%%%%%%%%%%%%%%%%%%%%%%%%%%%%%%%%%%%%%%%%%%%%%%%%%%%%%%%%%%%%
%
A \emph{\rpo} is periodic in its mean velocity
$\velRel_p=\gSpace_p/\period{p}$ co-rotating frame,
\reffig{f:MeanVelocityFrame}, but in the stationary frame its
trajectory is quasiperiodic. A co-moving frame is helpful in
visualizing a single `relative' orbit, but useless for
viewing collections of orbits, as each one drifts with its
own group velocity. A simultaneous visualization of all \rpo
s as \po s {can  be attained} only by \emph{symmetry
reduction}, to be undertaken in \refsects{s:Hilbert}{sec:mf}.

\Reqva\ and \rpo s are the hallmark of systems with
continuous symmetry. Amusingly, in this extension of periodic
orbit theory from unstable 1-dim\-ens\-ion\-al closed
periodic orbits to unstable $(N\!+\!1)$-dim\-ens\-ion\-al
compact manifolds $\pS_p$ in\-vari\-ant under continuous
symmetries, there are either no or proportionally few
periodic orbits. In presence of a continuous and no discrete
symmetry, likelihood of finding a {\po} is {\em zero}. \Rpo s
are almost never eventually periodic, \ie, they almost never
lie on periodic trajectories in the full {\statesp}, so
looking for periodic orbits in systems with only continuous
symmetries is a fool's errand.

\emph{A historical note.}
\Reqva\ and \rpo s are
related to \eqva\ and \po s of dynamics
reduced by the symmetries. They appear in many physical
situations, such as motion of rigid bodies, gravitational
$N$-body problems, molecules, nonlinear waves, spiralling
patterns and turbulence. According to Cushman,
Bates\rf{CushBat97} and Yoder\rf{Yode88}, C.
Huygens\rf{Huyg1673} understood the \reqva\ of a spherical
pendulum many years before publishing them in 1673. A
reduction of the translation symmetry was obtained by Jacobi
(for a modern, symplectic implementation, see Laskar
\etal\rf{MaRoLa02}). According to Chenciner\rf{Chenc05}, the
first attempt to find (relative) periodic solutions of the
$N$-body problem was the 1896 short note by
Poincar\'e\rf{Poinc1896}, in the context of the 3-body
problem. \Reqva\ of the $N$-body problem (known in this
context as Lagrange points, stationary in the co-rotating
frame) are circular motions in the inertial frame, and {\rpo
s} correspond to quasiperiodic motions in the inertial frame.
\Reqva\ that exist in a rotating frame are called central
configurations. For \rpo s in celestial mechanics see also
\refref{Broucke75}. A striking application of \rpo s has been
the discovery of `choreographies' of $N$-body problems%
\rf{CheMon00,CGMS02,McCordMontaldi}.

The modern story on equivariance and dynamical systems starts
perhaps with M. Field\rf{Field70}, and on bifurcations in
presence of symmetries  with Ruelle\rf{ruell73}. Ruelle
proves that the \stabmat/\jacobianM\ evaluated at an
\eqv/fixed point $\ssp \in \pS_G$ decomposes into linear
irreducible representations of \Group, and that
stable/unstable manifold continuations of its eigenvectors
inherit their symmetry properties, and shows that an \eqv\
can bifurcate to a rotationally in\-vari\-ant periodic orbit
(\ie, \reqv).

\subsection{\label{s:CLEsols} An example: Solutions of the \cLe}

In the case of the {\cLe}  the origin \EQV{0} is an \eqv\ of
\refeq{eq:CLe} for any value of the parameters. It is stable
for $0<\RerCLor<\rho_{1c}$ and unstable for
$\rho_{1c}<\RerCLor$, where\rf{FowlerCLE82}
\[
	\rho_{1c} = 1 + {(e+\ImrCLor)(e-\sigma \ImrCLor)}/{(\sigma+1)^2}
\,.
\]
At the bifurcation\rf{ruell73} a pair of eigenvalues crosses
the imaginary axis with imaginary part
\beq
	\omega_c = {\sigma (e + \ImrCLor)}/{(\sigma+1)}
\,,
\ee{eq:omegaCLE}
and a \emph{relative equilibrium} \REQV{}{1} with constant
angular velocity $\omega_c$ is born. For $\omega_c =0$ the
\reqv\ degenerates to an \SOn{2}-orbit of \eqva. As the
existence of a \reqv\ in a system with \SOn{2} symmetry is
the generic situation, we follow \refref{BakasovAbraham93}
and set $\ImrCLor=0$ and $e \neq 0$.

To find the location of the \reqv\ it is convenient to work
in polar coordinates
\beq
(x_1,x_2,y_1,y_2,z) =
    (r_1 \cos\theta_1,r_1\sin\theta_1,
     r_2\cos\theta_2,r_2\sin\theta_2,z)
\,,
\label{eq:CartToPol}
\eeq
where $r_1 \geq 0 \,,r_2 \geq 0$.
The \cLe\ \refeq{eq:CLe} take the form
\[
\left(
\begin{array}{c}
\dot{r}_1\\
\dot{\theta}_1\\
\dot{r}_2\\
\dot{\theta}_2\\
\dot{z}
\end{array}
\right)
=
\left(
\begin{array}{c}
 -\sigma\left(r_1 - r_2\cos\theta\right) \\
 -\sigma\frac{r_2}{r_1}\sin \theta  \\
 -r_2 + r_1\left((\rho_1-z)\cos \theta - \rho_2 \sin\theta\right)\\
  e  + \frac{r_1}{r_2}\left((\rho_1-z)\sin\theta +\rho_2 \cos\theta\right)\\
 -b z + r_1 r_2\cos\theta
\end{array}
\right)
,
\]
For
rotationally in\-vari\-ant flows the dynamics depends only
on the relative angle $\theta = \theta_1-\theta_2$
(which is why one speaks of `relative' equilibria).
This observation enables us to recast the \cLe\
in the  4-dimensional \reducedsp:
\beq
\left(
\begin{array}{c}
\dot{r}_1\\
\dot{r}_2\\
\dot{\theta}\\
\dot{z}
\end{array}
\right)
=
\left(
\begin{array}{c}
 -\sigma\left(r_1 - r_2\cos\theta\right) \\
 -r_2 + (\rho_1-z)r_1\cos \theta\\
  -e -\left(\sigma\frac{r_2}{r_1}
 +(\rho_1-z)\frac{r_1}{r_2}\right)\sin\theta\\
 -b z + r_1 r_2\cos\theta
\end{array}
\right)
\,,
\label{eq:PolarCLeTheta}
\eeq
where we have set $\rho_2=0$. The full 5-dimensional
evolution can be regained by integrating the driven
{\em
reconstruction} equation for the mean angular velocity:
\beq
\dot{\theta}_1 + \dot{\theta}_2
=
  e -\left[\sigma\,{r_2}/{r_1}
           -(\rho_1-z)\,{r_1}/{r_2}\right]\sin\theta
\,.
\label{eq:PolarCLeAngles}
\eeq
In general $\theta_1$ and
$\theta_2$ change in time, but for the \reqva\ the
difference between them is constant.
The condition for a \reqv\ is that all
time derivatives in \refeq{eq:PolarCLeTheta} vanish, while
$\dot{\theta}_1=\dot{\theta}_2\neq 0$ (if
$\dot{\theta}_1=\dot{\theta}_2=0$ we have a group orbit
of \eqva\ instead).
The \reqv\
$\REQV{}{1}$ is given by
\bea
(r_1,r_2,\theta,z) &=&
\left(\sqrt{b \,(\rho_1-d)},  \sqrt{b d \,({\rho_1}-d}),
     \cos^{-1}({1}/{\sqrt{d}}),  \rho_1-d
\right)
\,,
\label{eq:E1-PC}
\eea
where $d=1 + {e^2}/{(\sigma +1)^2}$, and
its angular velocity is
\beq
\dot{\theta}_{i}
= {\sigma e}/{(\sigma + 1)}
\,,
\label{eq:REQV1veloc}
\eeq
with period
$\period{{\REQV{}1}}= 2\pi (\sigma + 1)/\sigma e$.
For the parameter values \refeq{eq:CLeR}, the \reqv\ is at
\beq
\ssp_{\REQV{}1} = (r_1,r_2,\theta,z) =
     (8.48527,
      8.48562,
      0.00909,
      26.9999)
\,,
\label{eq:Q1}
\eeq
rotating with the period $\period{{\REQV{}1}}=69.1150$.

As $\RerCLor$ is increased,  a secondary bifurcation from
\REQV{}{1} results in a \emph{\rpo} \refeq{RPOrelper1}, or,
more precisely, in the quasiperiodic 2-frequency
\emph{modulated traveling wave}\rf{Krupa90}. With further
increase in $\RerCLor$ the dynamics turns chaotic, with {an}
infinity of unstable {\rpo s}. Once symmetry reduced maps are
constructed (see \reffig{fig:CLEip}\,(b) below), a large number of
these can be computed by methods described
elsewhere\rf{SCD07,SiminosThesis}. Calculation of {the}
\REQV{}{1} stability eigenvalues for the parameter values
\refeq{eq:CLeR}
(see \refref{SiminosThesis} for a calculation of stability of
\reqva\ in equi\-vari\-ant variables)
yields a weakly unstable spiral-out
\eqv\
\beq
(\eigExp_{1,2},\eigExp_3,\eigExp_4)
= (0.0938179 \pm 10.1945 i,-11.0009,-13.8534)
\,.
\ee{eq:CLeREQBstab}

The role of {the} above exact in\-vari\-ant solutions is illustrated by the
portrait of \cLf\ \statesp\ in  \reffig{fig:CLE}, with the
\reqv\ \REQV{}{1} and three repetitions of {the} \cycle{01} \rpo\
superimposed over a generic chaotic orbit. Repeats of
\cycle{01} {trace out a torus ergodically}, so in a system with
a $1$-dimensional continuous symmetry the organizational
blocks of a strange attractor are circles (\reqva) instead of
points (\eqva), and partially hyperbolic tori (\rpo s)
instead of closed loops (\po s). It is difficult to
understand the geometry of the flow by looking at such tori.

The large imaginary part of $\eigExp_{1}$ in
\refeq{eq:CLeREQBstab} implies that the simulation has to be
run up to time of order of at least 70 for the strange
attractor in \reffig{fig:CLE} to start filling in. Dynamics
is organized by the interplay of the stable and unstable
manifolds of \eqv\ \EQV{0} and \reqv\ \REQV{}{1}, but the
symmetry-induced drift along the direction of rotation blurs
the picture and the notion of recurrence becomes relative. In
what follows, it is this confusing situation (as well as the
theoretical fact\rf{Cvi07} that dynamical zeta functions have
their support on \rpo s) that motivates the search for
effective methods to project the dynamics onto a \reducedsp.

\section{\label{s:symmRed} Symmetry reduction}

The action of a symmetry group \Group\ on \pS\ endows the
\statesp\ with the structure of a union of group orbits, each
group orbit an equivalence class. The goal of {\em symmetry
reduction} is the identification of a unique point as the
representative of a group orbit, and the replacement of the
original \statesp\ by the space of such points, the {\em
\reducedsp}. In the literature this space is alternatively
called
\emph{desymmetrized \statesp},
\emph{symmetry-reduced space},
\emph{orbit space}, or \emph{quotient space}
$\pS/\Group$ because symmetry has been `divided out.' {The}
symmetry group \Group\ of equi\-vari\-ant dynamics acts
trivially in {the} \reducedsp, and the resulting dynamical
system, called by Gilmore and Lettelier\rf{GL-Gil07b} the
\emph{image}, is symmetry {\em in\-vari\-ant}, in the sense
that its symmetry group is the identity. \Reducedsp\ is in
general not a manifold but rather a union of manifolds of
different dimensions\rf{ChossLaut00}.

In \refsect{s:Hilbert} we briefly review one of the standard
tools by which spatial symmetry reduction can be achieved:
projection to a \emph{Hilbert basis}, and show in
\refsect{s:cLeHilbert} how it works for \cLf. A wonderful
symmetry reduction tool for low-dimensional flows, the
Hilbert basis approach turns out to be too cumbersome to be
applicable to high-dimensional flows. Next we describe the
\emph{\mframes} (\refsect{sec:mf}) and apply it to the \cLf\
example to illustrate the form of a general linear slice
(\refsect{sec:CLeMovFr}), show how the method enables us
to explicitly compute \Group-in\-vari\-ant coordinates, and
relate these to the Hilbert in\-vari\-ant polynomial basis
(\refsect{s:cleCoordSlice}). Then we discus different choices
of slice-fixing points (\refsect{s:mfReqb}), and the
associated singular sets.
Since rotations commute with time integration, one can
start with a point on the \slice,
integrate for short time and then rotate the
end point back into the \slice.
In \refsect{sec:MovFrameODE} the limit of
infinitesimal time steps yields the equivalent but differential
formulation, the \emph{\mslices} for which the flow is restricted to
the \reducedsp.
In practice we find it more convenient to use the
numerical code as given, and post-process the data by the \mframes,
rather than rewriting the equations in the \mslices\ form.

\section{\label{s:Hilbert} Hilbert polynomial bases}

In atomic physics and other low-dimensional physical problems
with spatial symmetries, symmetry reduction is customarily
implemented just as we did it in \refeq{eq:CartToPol}, by
going to the natural coordinate system (polar, cylindrical,
\etc). That works well for linear systems, but not so well
for nonlinear flows, and some take pride in
using no polar coordinates in symmetry reduction of
Hamiltonian flows\rf{SadoEfst05,CushBat97}; note, for
example, that these coordinate transformations introduce
singularities in \refeq{eq:PolarCLeTheta} at $r_1=r_2=0$.

What are we really doing when redefining dynamics in terms of
such in\-vari\-ant coordinates? We are recasting
equi\-vari\-ant dynamics of $(\ssp_1,\ssp_2,\cdots)$
coordinates in terms of rotationally in\-vari\-ant lengths
$(r_1=(\ssp_1^2+\ssp_2^2)^2,\cdots)$, volumes and other
in\-vari\-ant quantities. Physical laws have the same form in
all coordinate frames, so they are often formulated in terms
of functions (Hamiltonians, Lagrangians, $\cdots$) {that are}
in\-vari\-ant under a given set of symmetries. Given a
symmetry, what is the most general functional form of such
law? The general problem of symmetry reduction in this sense
{was} elegantly solved nearly a century ago. According to the
Hilbert-Weyl theorem, for a compact group $\Group$ there
exists a finite $\Group${-in\-vari\-ant} Hilbert polynomial
basis $\{u_1,u_2, \dots,u_m\}$, $ m \geq d$, such that any
$\Group${-in\-vari\-ant} polynomial can be written as a
multinomial
\beq
h(\ssp) = p(u_1(\ssp),u_2(\ssp), \dots,u_m(\ssp))
    \,,\qquad \ssp \in \pS
\,.
\ee{HilbWeyl}
The Gilmore and Lettelier monograph\rf{GL-Gil07b} offers a
clear, detailed and user friendly discussion of symmetry
reduction by means of in\-vari\-ant polynomial bases (do not
look for Hilbert in the index, though). The in\-vari\-ant
dynamical equations follow from the equi\-vari\-ant ones by
chain rule
\beq
 \dot{ u}_i=\frac{\partial u_i}{\partial x_j} \, \dot{x}_j
 \,,
\ee{HilbChainRl}
upon substitution $\{x_1,x_2,\cdots,x_d\}$ $\to$
$\{u_1,u_2,\cdots,u_m\}$. One can either rewrite the dynamics
in this basis, or one can simply plot the `image' of
solutions computed in the original, equi\-vari\-ant basis in
terms of these in\-vari\-ant polynomials.

Unfortunately, while the idea is elegant, an explicit
construction of $\Group${-in\-vari\-ant} basis can in
practice be a daunting undertaking. The set  of $m \geq d$
in\-vari\-ant polynomials $\{u_1,u_2,\cdots,u_m\}$ is not
unique, and while these polynomials are linearly independent,
they are functionally dependent through $m-d+N$ nonlinear
relations called \emph{syzygies}. Their determination becomes
quickly computationally prohibitive as the dimension of the
system and/or group increases\rf{gatermannHab,ChossLaut00},
and in practice computations are confined to
dimensions less than ten. As our goal is to quotient
continuous symmetries of high-dimensional flows, as high as
$10^2$-$10^6$ coupled ODEs arising from
truncations of the \KS\ and Navier-Stokes flows, reduction by
the method of Hilbert basis is at present not a feasible
option.

Nevertheless, as symmetry reduction of moderate-dimension
flows by the method of in\-vari\-ant polynomials offers a clean
benchmark for other approaches to symmetry reduction, we start
by showing how it works for \cLf.

\subsection{\label{s:cLeHilbert} An example: \CLe\ recast in Hilbert basis}

\begin{figure}[ht]
\begin{center}
(\textit{a})\includegraphics[width=0.35\textwidth,clip=true]{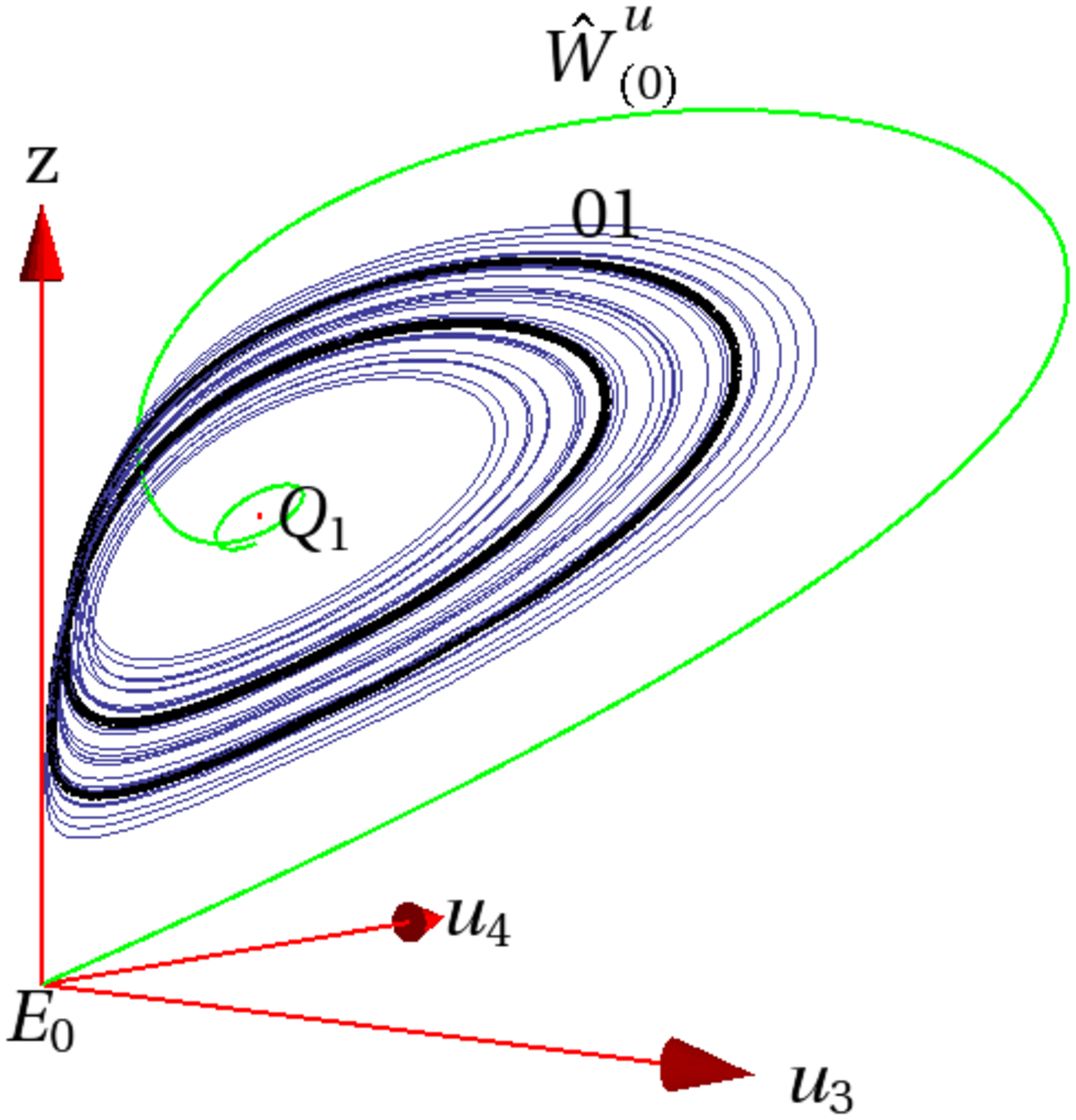}
 ~~~~(\textit{b})\includegraphics[width=0.35\textwidth,clip=true]{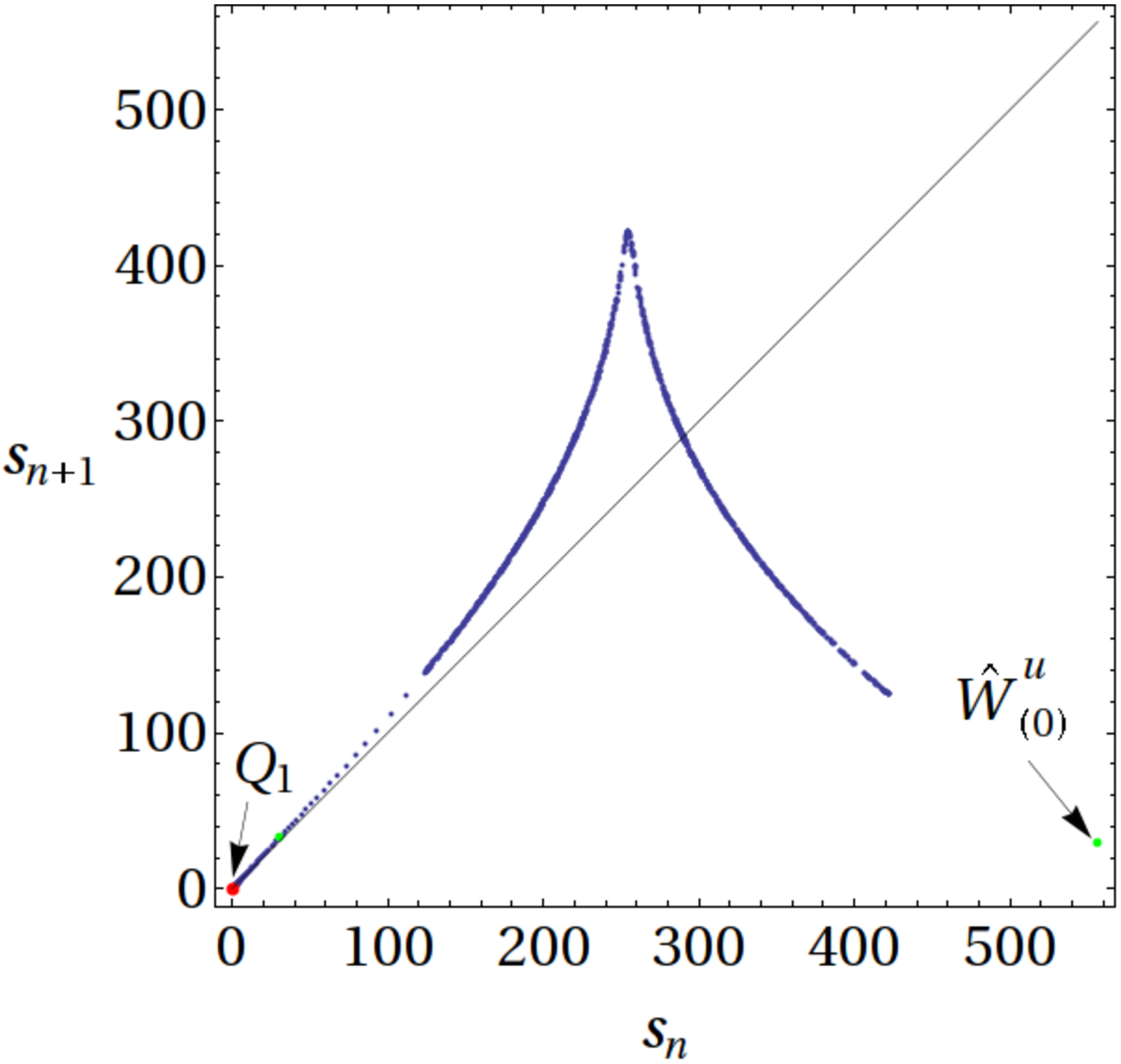}
\end{center}
\caption{
(a) Invariant image of \cLf, \reffig{fig:CLE}, projected onto
the in\-vari\-ant polynomials basis \refeq{eq:ipLaser}. Note
the unstable manifold connection from the \eqv\ $\EQV{0}$ at
the origin to the strange attractor controlled by the
rotation around the \reducedsp\ image of \reqv\ \REQV{}{1};
as for the Lorenz flow\rf{lorenz}, the natural measure close
to $\EQV{0}$ is vanishingly small but non-zero.
(b) The return map for the \Poincare\ surface of section
$u_1=u_4$ for the \cLe\ projected on in\-vari\-ant polynomials
\refeq{eq:ipLaser}. The return map coordinate is the
Euclidean length \refeq{EuclArcl} along the \Poincare\
section of the unstable manifold of $\REQB{1}$.
    }
\label{fig:CLEip}
\end{figure}

As the Hilbert basis approach turns out to be too cumbersome
for our main goal, symmetry reduction of high-dimensional
flows, we forgo here a  systematic discussion of how to
construct in\-vari\-ant polynomials bases. The pedagogical
literature mostly focuses on discrete symmetry
groups\rf{GL-Gil07b,gatermannHab,ChossLaut00}, while general
algorithms are a domain of advanced algebraic geometry
monographs. For the purpose at hand it suffices to use the
Gilmore and Letellier\rf{GL-Gil07b,Letellier03} in\-vari\-ant
polynomial basis for the action \refeq{CLfRots}. They apply
it to the symmetry reduction of the Zeghlache-Mandel
system\rf{ZeMa85}, a flow much like the \cLe. As it can be
easily verified, the Hilbert basis
\bea
        u_1 &=& x_1^2+x_2^2\,,\qquad\qquad
        u_2 \,=\, y_1^2+y_2^2 \continue
        u_3 &=& x_1 y_2-x_2 y_1\,,\qquad
        u_4 \,=\, x_1 y_1+x_2 y_2	\label{eq:ipLaser}\\
        u_5 &=& z
\nnu
\eea
is in\-vari\-ant under \refeq{CLfRots}, the $\SOn{2}$ action on a
5-dim\-ens\-ion\-al \statesp. That implies,
in particular, that the image of the full \statesp\ \reqv\
\REQV{}{1} group orbit of \reffig{fig:CLE} is the
\eqv\ point in \reffig{fig:CLEip}\,(a),
while the image of a \rpo, such as \cycle{01}, is a
periodic orbit. The
five polynomials are linearly independent, but related through one
syzygy,
\beq
u_1 u_2 -u_3^2-u_4^2 =0
  \,,
\label{eq:syzLaser}
\eeq
yielding a 4-dim\-ens\-ion\-al $\pS/\SOn{2}$ \reducedsp,
a symmetry-in\-vari\-ant representation of the 5-dim\-ens\-ion\-al
\SOn{2} equi\-vari\-ant dynamics.
%
%%%%%%%%%%%%%%%%%%%%%%%%%%%%%%%%%%%%%%%%%%%%%%%%%%%%%%%%%%%%%%%%%%
%%%%%%%%%%%%%%%%%%%%%%%%%%%%%%%%%%%%%%%%%%%%%%%%%%%%%%%%%%%%%%%%
Vladimirov, Toronov and Derbov\rf{VlToDe98}
use a different invariant polynomial basis to study bounding
manifolds of the symmetry reduced \cLf\ and its homoclinic bifurcations.

Application of the chain rule \refeq{HilbChainRl} brings the
equi\-vari\-ant \cLe\ \refeq{eq:CLeR} to the in\-vari\-ant form
\refeq{eq:ipLaser}:
\bea
\dot{u}_1 &=&2\,\sigma\,(u_4-u_1)\,,
\continue
\dot{u}_2 &=&-2\left(\,u_2 - \rho_2\, u_3 -\,(\rho_1-u_5)\,u_4\right)\,,\continue
\dot{u}_3 &=& -(\sigma\, +1)\,u_3+\rho_2\, u_1+e\, u_4\,,    \label{eq:CLEip}\\
\dot{u}_4 &=& -(\sigma\, +1)\,u_4+\,(\rho_1-u_5)\,u_1+\sigma\, u_2-e\,u_3\,,\continue
\dot{u}_5 &=& u_4-b\, u_5\,.
\nnu
\eea
As far as visualization goes, we need neither construct the
in\-vari\-ant equations \refeq{eq:CLEip} nor integrate them. It
suffices to integrate the original, unreduced flow of
\refFig{fig:CLE}, but plot the solution in the image space,
\ie, the in\-vari\-ant, Hilbert polynomial coordinates $u_i$, as
in \reffig{fig:CLEip}\,(a). A minor drawback of the Hilbert
polynomial basis projections is that the folding mechanism is
harder to view since the dynamics is squeezed near the
$z$-axis.

\subsection{\label{s:Poincare}Symmetry reduced return map}

Successive trajectory intersections with a {\em Poincar\'e
section}, a $(d-1)$-dim\-ens\-ion\-al manifold or a set of
manifolds $\PoincS$ embedded in the $d$-dim\-ens\-ion\-al
{\statesp} $\pS$, define the {\em Poincar\'e return map}
$\PoincM({\ssp})$, a $(d-1)$-dim\-ens\-ion\-al map of form
\beq
\ssp' = \PoincM({\ssp})
          =  \flow{\tau(\ssp)}{\ssp}
\,,\qquad
\ssp', \ssp \in \PoincS
\,.
\ee{PoincMap}
Here the {\em first return function} $\tau(\ssp)$, sometimes
referred to as the {\em ceiling function}, is the time of
flight to the next {Poincar\'e} section for a trajectory starting at
$\ssp$. A good choice of the {Poincar\'e} section manifold $\PoincS$
in high-dimensional ($d>3$) flows
is basically a dark art. We chose a Poincar\'e section which contains the $z$-axis and
the \reqv, here defined by the condition $u_1=u_4$.
Even though in the \cLe\ case we could use one of the $u_i$'s
as a coordinate to construct a return map, for high dimensional
flows a dynamically intrinsic parametrization is the only option.
Following \refref{Christiansen97}, we construct the first
return map of \reffig{fig:CLEip}\,(b) using as a coordinate
the Euclidean length along the intersection of the unstable
manifold of \REQV{}{1} within the Poincar\'e surface of
section, measured from \REQV{}{1}.

We begin by sprinkling evenly spaced points
$\{\ssp^{(1)},\ssp^{(2)},\cdots,\ssp^{(N-1)}\}$ between the
\reqv\ point $\ssp_{\REQV{}{1}}=\ssp^{(0)}$ and the point
$\ssp= \ssp^{(N)}$, along the $1d$ intersection of \Poincare\
section and unstable manifold continuation $\ssp^{(k)} \in
\hat{W}^u_{(1)}$ of the unstable ${\bf \hat{e}}^{(1)}$
eigen\-plane (we shall omit the eigen\-direction label
${}_{(1)}$ in what follows). Then the arclength in Euclidean
metric from the
\reqv\ point $\ssp_{\REQV{}{1}}=\ssp^{(0)}$ to $\ssp=
\ssp^{(N)}$ is given by
\beq
s = \lim_{N\to\infty}
         \sqrt{
            \sum_{k=1}^N \left(d\ssp^{(k)}\right)^2
              }
\,.
\ee{EuclArcl}
By definition $f^{\tau(\ssp)}(\ssp) \in \hat{W}^u_{(j)}$, so
$f^t(\ssp)$ induces a $1d$ map $s(s_0,\tau) =
s(f^{\tau(\xInit)}(\xInit))$.

{\em \Turn s} are points on the unstable manifold for which
the local unstable manifold curvature diverges for forward
iterates of the map, \ie, points at which the manifold folds
back onto itself arbitrarily sharply. The $1d$ curve $\hat{W}^u_{(1)}$
starts out linear at $\ssp_{\REQV{}{1}}$, then gently curves
until it folds back sharply at the `\turn' along a possible
heteroclinic connection to \EQV{0}, and then nearly retraces
itself.

The trick is to figure out a good {\em base segment} to the
nearest {\turn} $L=[0,s_{b}]$, and after the foldback assign
to $s(\ssp,t)>s_{b}$ the nearest point $s$ on the base
segment. Since here the stable manifold contraction is strong, the
2nd coordinate connecting $s(\ssp,t) \to s$ can be neglected.

Armed with this intrinsic curvilinear coordinate
parametrization, we are now in a position to construct a
1-dimensional model of the dynamics on the \nws. If
$\hat{\ssp}_n$ is the $n$th Poincar\'e section of a
trajectory in neighborhood of $\ssp_\stagn$, and $s_n$ is the
corresponding curvilinear coordinate, then $s_{n+1} =
f^{\tau_n}(s_n)$ models the full \statesp\ dynamics
$\hat{\ssp}_n \to \hat{\ssp}_{n+1}$. We approximate $f(s_n)$
by a smooth, continuous 1-dimensional map $f : L_\stagn \to
L_\stagn$ by taking $\hat{\ssp}_n \in L_\stagn$, and
assigning to $\hat{\ssp}_{n+1}$ the nearest base segment
point $s_{n+1}=s(\hat{\ssp}_{n+1})$. Thanks to the extreme
contraction rate \refeq{trA-ZM}, the return map turns out to
be unimodal for all practical purposes, so binary symbolic dynamics are easily
constructed and admissible \po s of the map up to desired length can be
systematically obtained. A multiple shooting routine\rf{DasBuch} can then be
used to determine the corresponding \rpo s
of the \cLe\ to machine precision.

\section{\label{sec:mf} The method of moving frames} 

The \mframes, introduced by G. Darboux and systematized by
\'E. Cartan\rf{CartanMF}, is interpreted by Fels and
Olver\rf{FelsOlver98,FelsOlver99} as a map from a
$d$-dimensional manifold
to a $N$-dimensional Lie group acting on it.
Moving frames are then used to
compute $d-N$ functionally independent \emph{fundamental
in\-vari\-ants} for general group actions in relation to general
equivalence problems. `Fundamental' here means that they can
be used to generate all other in\-vari\-ants, and, in particular,
they serve to distinguish group orbits in an open
neighborhood of the {\slice} point, \ie, two points lie on
the same group orbit if and only if all fundamental
in\-vari\-ants agree. For an introduction to the method we
recommend Olver's pedagogical monograph\rf{OlverInv}. Here we
emphasize the application of the method to dynamical symmetry
reduction, and focus in particular on groups
acting on spaces with the structure of a direct sum of
irreducible subspaces, an application which is not to our knowledge
explored in the literature. We are not
concerned with the explicit
determination of the fundamental in\-vari\-ants as in
\refref{FelsOlver98,FelsOlver99}, except for illustrative
purposes; instead we focus on
implementation of a moving frame transformation as a
numerically fast and efficient linear map of the full
\statesp\ dynamics onto its desymmetrized, \reducedsp\
projection.

The main idea behind \mframes\ is that we can, at least locally,
map each point along any solution $\ssp(\tau)$ to a unique
representative $\sspRed(\tau)$ of the associated
group orbit equivalence class, by a suitable rotation
\beq
\ssp(\tau) = \LieEl(\tau) \, \sspRed(\tau)
\,.
\ee{EquiTraj}
Equivariance implies the two points are equivalent. In the
\mframes\ the \reducedsp\ representative $\sspRed$ of a
group orbit equivalence class is picked by slicing across the
group orbits by a fixed manifold.
%
%%%%%%%%%%%%%%%%%%%%%%%%%%%%%%%%%%%%%%%%%%%%%%%%%
\begin{figure}[ht]
\begin{center}
(\textit{a})\includegraphics[width=0.48\textwidth,clip=true]{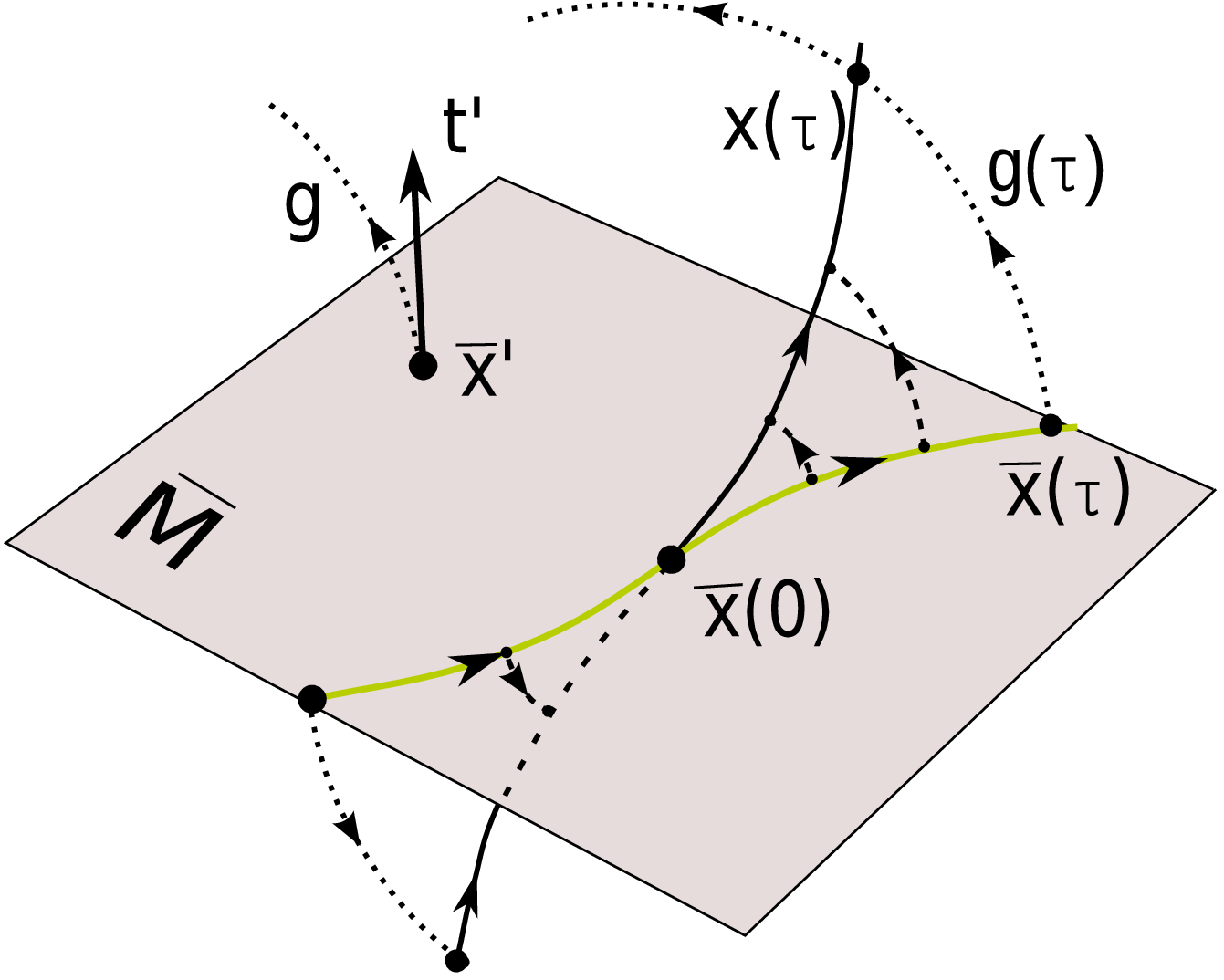}
~~~~(\textit{b})\includegraphics[width=0.33\textwidth,clip=true]{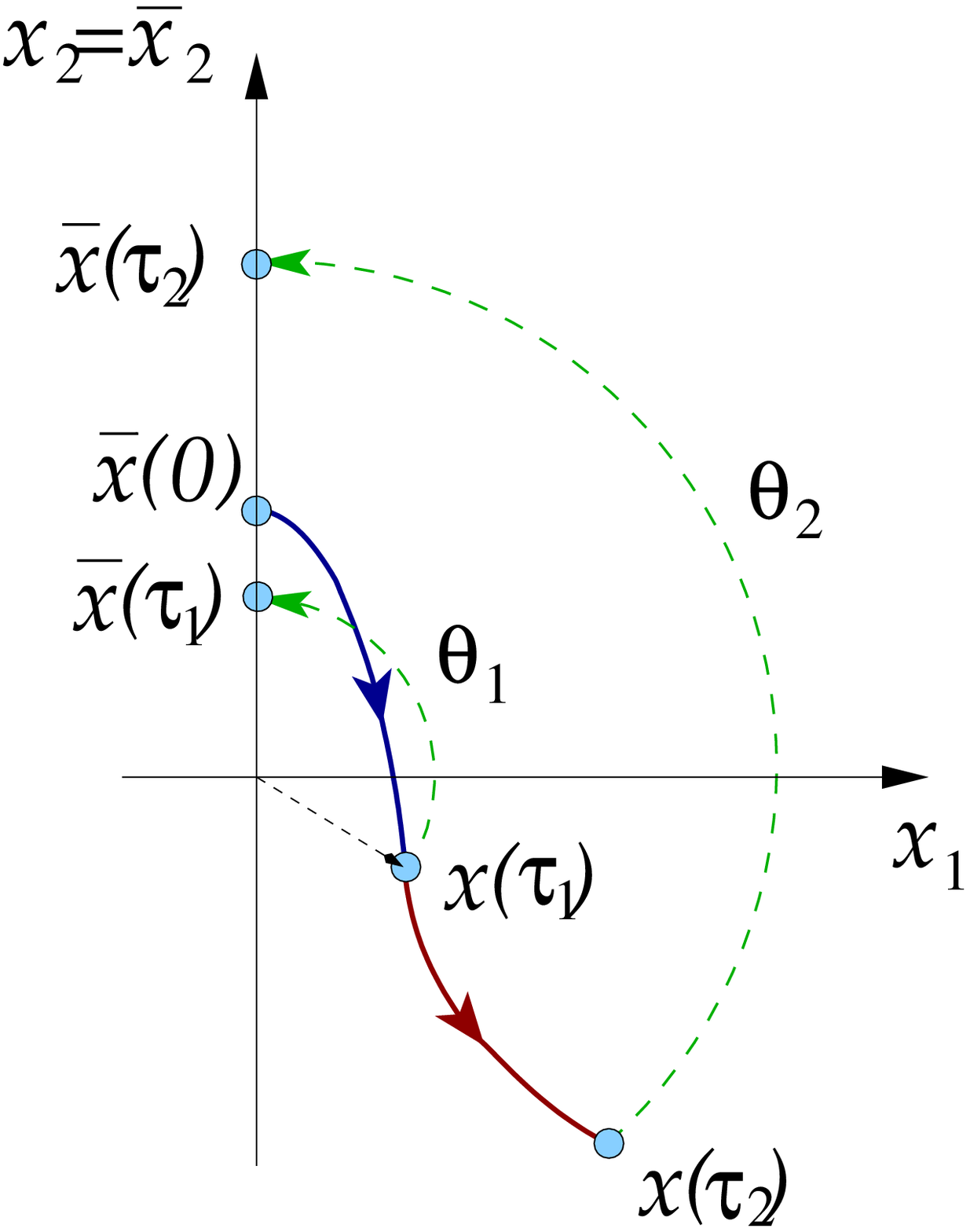}
\end{center}
\caption{
(a)
\Slice\ \pSRed\ is a hyperplane \refeq{PCsectQ}
passing through the slice-fixing point $\slicep$,
and normal to the group tangent $\sliceTan{}$ at $\slicep$.
It intersects all
group orbits (indicated by dotted lines here) in an open
neighborhood of $\slicep$.  The full
\statesp\ trajectory $\ssp(\tau)$ and the \reducedsp\
trajectory $\sspRed(\tau)$ belong to the same group orbit
$\pS_{\ssp(\tau)}$ and are equivalent up to a group rotation
$\LieEl(\tau)$.
~(b)
The \mframes\ for a flow $\SOn{2}$-equi\-vari\-ant under
\refeq{CLfRots} with \slice\ through $\slicep=(0,1,0,0,0)$,
group tangent $\sliceTan{}=(-1,0,0,0,0)$. The
orientation condition restricts the \slice\ to half-hyperplane
$\sspRed_1=0,\;\sspRed_2\ge 0$. A trajectory started on the
\slice\ at $\sspRed(0)$ evolves to a \statesp\ point with a
non-zero $\ssp_1(\tau_1)$. Compute angle $\gSpace_1$
through slice condition \refeq{PCsectQ1}. Rotate $\ssp(\tau_1)$
counter-clockwise by $\gSpace_1$ to $\sspRed(\tau_1) =
\LieEl(\gSpace_1)\,\ssp(\tau_1)$, so that the equivalent point
on the circle lies on the \slice, $\sspRed_1(\tau_1) =0$. Repeat
for all sample points $\ssp(\tau_i)$ along the trajectory.
}\label{fig:ReducTraj}
\end{figure}
%%%%%%%%%%%%%%%%%%%%%%%%%%%%%%%%%%%%%%%%%%%%%%%%%%
%

In the following it will be useful to introduce the notion of
a \emph{\slice}, an $(d-N)$-dimensional submanifold
$\pSRed\subset\pS$ such that $\pSRed$ intersects all group
orbits in an open neighborhood of $\slicep \in \pSRed$
transversally and at most once. In other words, \slice\ is
the analogue of a Poincar\'e section, but for group orbits.
As is the case for the dynamical Poincar\'e sections, in
general a single \slice\ does not suffice to intersect all
group orbits of points in \pS. One can construct a local {\slice} passing
through any point $x\in \pS$ if the group orbits of \Group\
have the same dimension, \ie\ {they are} away from {\fixedsp
s} of continuous subgroups of \Group, see
\refref{FelsOlver99} for details.

The simplest {\em \slice\ condition} defines a linear \slice\ as a
$(d\!-\!N)$-dim\-ens\-ion\-al hyperplane \pSRed\ normal to
the $N$ group rotation tangents $\sliceTan{a}$ at point $\slicep$,
see \reffig{fig:ReducTraj}:
\beq
(\sspRed - \slicep )^T \sliceTan{a} =0
    \,,\qquad
\sliceTan{a} = \groupTan_a(\slicep) = \Lg_a \, \slicep
\,,\qquad a=1,2, \cdots, N\,.
\ee{PCsectQ}
As $\slicep^T \sliceTan{a} =0$ by the antisymmetry of
$\Lg_a$, the \slice\ condition \refeq{PCsectQ} fixes
$\gSpace$ for a given $\ssp$ by orthogonality,
\beq
0 = \sspRed^T  \sliceTan{a}
	=\ssp^T  \LieEl(\gSpace)^T \sliceTan{a}
\,,
\ee{PCsectQ1}
where $\LieEl^T$ denotes the transpose of $\LieEl$. A group
orbit will in general intersect a slice more than once (for
example in the case of $\SOn{2}$, two $\pi$-separated
points), so we need to impose further conditions on the
slice, in the form of either inequalities or orientation
conditions, so as to ensure unique intersection. These
restrictions are rather arbitrary, the only requirement being
that $\pSRed$ remains a connected manifold. We illustrate
this point for the \cLf\ example in \refsect{sec:CLeMovFr}.

For group orbits intersected by a slice, we can identify the
unique group element $\LieEl=\LieEl(\ssp)$ that rotates
$\ssp$ into the slice, $\LieEl\ssp = \sspRed \in \pSRed$. The
map that associates to a \statesp\ point $\ssp$ a Lie group
action $\LieEl(\ssp)$ is called a \emph{moving frame}.
The \mframes\ can be thought of as a change of variables
$
\sspRed = \LieEl^{-1}(x) \, \ssp
\,,
$ to a frame of reference for which the slice-fixing condition
\refeq{PCsectQ1} is identically satisfied--hence the name
`moving frame.'

The \mframes\ is a post-processing method; trajectories are
computed in the full \statesp, then rotated into the \slice\
whenever desired, with the \slice\ condition easily
implemented. The \slice\ group tangent \sliceTan\ \, is a
given vector, and rotation parameters $\gSpace$ are
determined numerically, by a Newton method, through the
\slice\ condition \refeq{PCsectQ1}. For given $\theta$,
$\LieEl(\gSpace)\,\ssp$ is another vector, linear in $\ssp$.

A \slice\ can be identified with $\pS/\Group$ in an open
neighborhood of $\slicep$. As is the case for the dynamical
Poincar\'e sections, a single \slice\ does not
suffice to reduce $\pS \to \pS/\Group$ globally as
 one cannot expect the group orbit
of any point in $\pS$ to intersect a given slice.

How does one pick a \slice\ point $\slicep$? A generic point
$\slicep $ not in an in\-vari\-ant subspace (on the $z$
axis of the \cLe, for example) should suffice to fix a \slice.
The rules of thumb are much like the ones for picking
Poincar\'e sections. The intuitive
idea is perhaps best visualized in the context of fluid
flows. Suppose the flow exhibits an unstable coherent
structure that {is frequently visited at}
different spatial dispositions. One can fit a `template' to one
recurrence of such structure, and describe other recurrences
as its translations. A well-chosen \slice\ point belongs to
{equivalence class that is dynamically important in
this way} (\ie, a group orbit).
We discuss, in the context of our \cLe\ example, several slice fixing
choices in \refsects{s:cleCoordSlice}{s:mfReqb}.

\emph{A historical note.}
For the definition of slice see, for example,  Chossat and
Lauterbach\rf{ChossLaut00}. Slices tend to be discussed in
contexts much more difficult than our application -
symplectic groups, sections in absence of global charts,
non-compact Lie groups. We follow
\refref{rowley_reconstruction_2000} in referring to a local
group-orbit section as a slice. The usage goes back at least
to Palais\rf{Pal61} in 1961 and Mastow\rf{Mostow57} in 1957.
Some\rf{FelsOlver99,Bredon72} refer to global group-orbit
sections as \emph{cross-sections}, a term that we would rather
avoid, as it has an honest, well-established meaning in
physics.
Guillemin and Sternberg\rf{GuiSte90} define the
cross-section, but emphasize that finding it is very
rare: ``existence of a global section is a very stringent
condition on a group action. The notion of slice is weaker
but has a much broader range of existence.''

\subsection{\label{sec:CLeMovFr}An example: Moving frame for \cLe}

In case of the \cLe\ we can, due to equivariance, rotate any
slice fixing point in \refeq{PCsectQ1} so that we have
$\slicepComp{x}{1}=0$.
As only the group tangent direction matters, a slice that
goes through point
    (0,\,\slicepComp{x}{2},\,\slicepComp{y}{1},\,
       \slicepComp{y}{2},\,\slicepComp{z}{})
is equivalent to
    (0,\,1,\,\slicepComp{y}{1}/\slicepComp{x}{2},\,
       \slicepComp{y}{2}/\slicepComp{x}{2},\,0),
so we can specify the most general slice fixing point for
the \cLe\ by two numbers,
\beq
\slicep=(0,\,1,\,\slicepComp{y}{1},\,\slicepComp{y}{2},\,0)
\,.
\ee{CLEslicFixP}
The group orbit tangent then becomes
$\sliceTan{}=(-1,\,0,\,-\slicepComp{y}{2},\,\slicepComp{y}{1},\,0)$
and slice condition \refeq{PCsectQ1} leads to
\beq
  \theta=\tan^{-1}\frac{x_1+\slicepComp{y}{2}y_1-\slicepComp{y}{1}y_2}
			  {x_2+\slicepComp{y}{1}y_1+\slicepComp{y}{2}y_2}\,.
\ee{cLeMF}
To ensure a unique intersection with the slice, we have to
further restrict $\pSRed$ by choosing a representative out of
the two group orbit points that intersect the slice. One can
impose an orientation condition, for example choosing the
point that is at minimum distance from $\slicep$, or one can
define the inverse tangent function $\tan^{-1}({b}/{a})$ so
that it distinguishes quadrants in the $(a,b)$ plane. Either
condition works equally well.

We observe that \refeq{cLeMF} is undefined when
\begin{subequations}\label{cLeMFsing}
  \begin{align}
    x_1+\slicepComp{y}{2}y_1-\slicepComp{y}{1}y_2 &=0 \label{cLeMFsingOnSlice}\cont
    x_2+\slicepComp{y}{1}y_1+\slicepComp{y}{2}y_2 &=0 \label{cLeMFsingPerpTan}\,,
  \end{align}
\end{subequations}
are both satisfied. We will refer to this $3$-dimensional
linear subspace as the \emph{\sset} of the moving frame
associated with \refeq{cLeMF}. Condition
\refeq{cLeMFsingOnSlice} implies that point \ssp\ is already
on the slice, $\ssp^T\slicep=0$. Condition
\refeq{cLeMFsingPerpTan} implies that the group tangent at
point \ssp\ is perpendicular to group tangent at slice fixing
point, $\groupTan{}(\ssp)^T\sliceTan{}=-\ssp^T\slicep=0$. The
problem lies in the fact that the limit of \refeq{cLeMFsing}
as we approach the \sset\ does not exist. For instance,
consider points for which \refeq{cLeMFsingPerpTan} holds; as
we approach the singularity with positive values of the
numerator in \refeq{cLeMF} we have $\theta=\pi/2$, while for
negative values $\theta=-\pi/2$. A $\pi$-jump occurs as we
cross the \sset. In general such crossings are expected to
occur, since the \sset\ is not flow in\-vari\-ant. Even
worse, the singularity distorts the way trajectories are
mapped onto the slice, even if they merely approach it rather
than cross it, as we will see in the next two examples.

\subsection{\label{s:cleCoordSlice}Irreducible subspace {\slice}
            and explicit in\-vari\-ants}

We now show how a particular choice of the slice point enables us
to express the transformation to in\-vari\-ant variables in a
simple analytic form.
Place the \slice\ point in one of the linearly irreducible
subspaces of $\SOn{2}$ action \refeq{CLfRots}, for instance
$\slicep=(0,-1,0,0,0)$. The group-tangent at the \slice\
point is then $\sliceTan{}=(1,0,0,0,0)$ and the slice fixing
condition is
\beq
    \overline{x}_1 = x_1 \cos\gSpace - x_2 \sin\gSpace = 0
\,.
\ee{cLeCoordSlice}
The orientation condition restricts the \slice\ to half-hyperplane
$\overline{x}_1=0,\;\overline{x}_2\ge 0$.
Solving \refeq{cLeCoordSlice}
for the polar angle $\gSpace$ in $(\ssp_1,\ssp_2)$ we get
\beq
  	\gSpace=\tan^{-1}({x_1}/{x_2})
\,.
\ee{cLeCoordTheta}
The transformation that rotates $\ssp$ counter-clockwise by $\gSpace$
to $\overline{\ssp} = \LieEl(\gSpace)\,\ssp$ onto the \slice\ is found by inserting
\refeq{cLeCoordTheta} into the expression for the action of \SOn{2}
on $\ssp$,
\bea
 	\overline{x}_1 &=& x_1 \cos\gSpace - x_2 \sin\gSpace
        \,,\quad
	\overline{x}_2  =  x_1 \sin\gSpace + x_2 \cos\gSpace
                    \label{eq:CLEexplSO2a}\\
	\overline{y}_1 &=& y_1 \cos\gSpace - y_2 \sin\gSpace
        \,,\quad
	\overline{y}_2 = y_1 \sin\gSpace + y_2 \cos\gSpace
                    \nnu
\eea
yielding the transformations in analytic form:
\bea
	\overline{x}_2 &=&  r_1 \,=\, \sqrt{x_1^2+x_2^2}
                \,,\qquad
    \overline{z} \,=\, z
                \continue
	\overline{y}_1 &=& {(x_2 y_1-x_1 y_2)}/{r_1}
                \,,\quad
	\overline{y}_2 \,=\, {(x_1 y_1+x_2 y_2)}/{r_1}
\,.
	\label{eq:invLaser}
\eea
This transformation rotates point $\ssp$ into the slice point
$\sspRed$. Alternatively, the transformation can be viewed as providing
in\-vari\-ant variables on which to project dynamics, as we did
in Hilbert basis case. Note the relation to the in\-vari\-ant
polynomials \refeq{eq:ipLaser}, and observe that as the
rotational degree of freedom has been explicitly  used, the
\mframes\ requires no syzygies.
Analytical determination of in\-vari\-ant variables through
moving frame transformations and particular slice conditions can
be carried out systematically
for general group representations\rf{FelsOlver98}, but
for reasons explained in \refsect{s:mfReqb}, we shall not
take this path here.

%
%%%%%%%%%%%%%%%%%%%%%%%%%%%%%%%%%%%%%%%%%%%%%%%%%%%%%%%%%%%%%%%%
\begin{figure}[ht]
\begin{center}
  (\textit{a})\includegraphics[width=0.34\textwidth,clip=true]{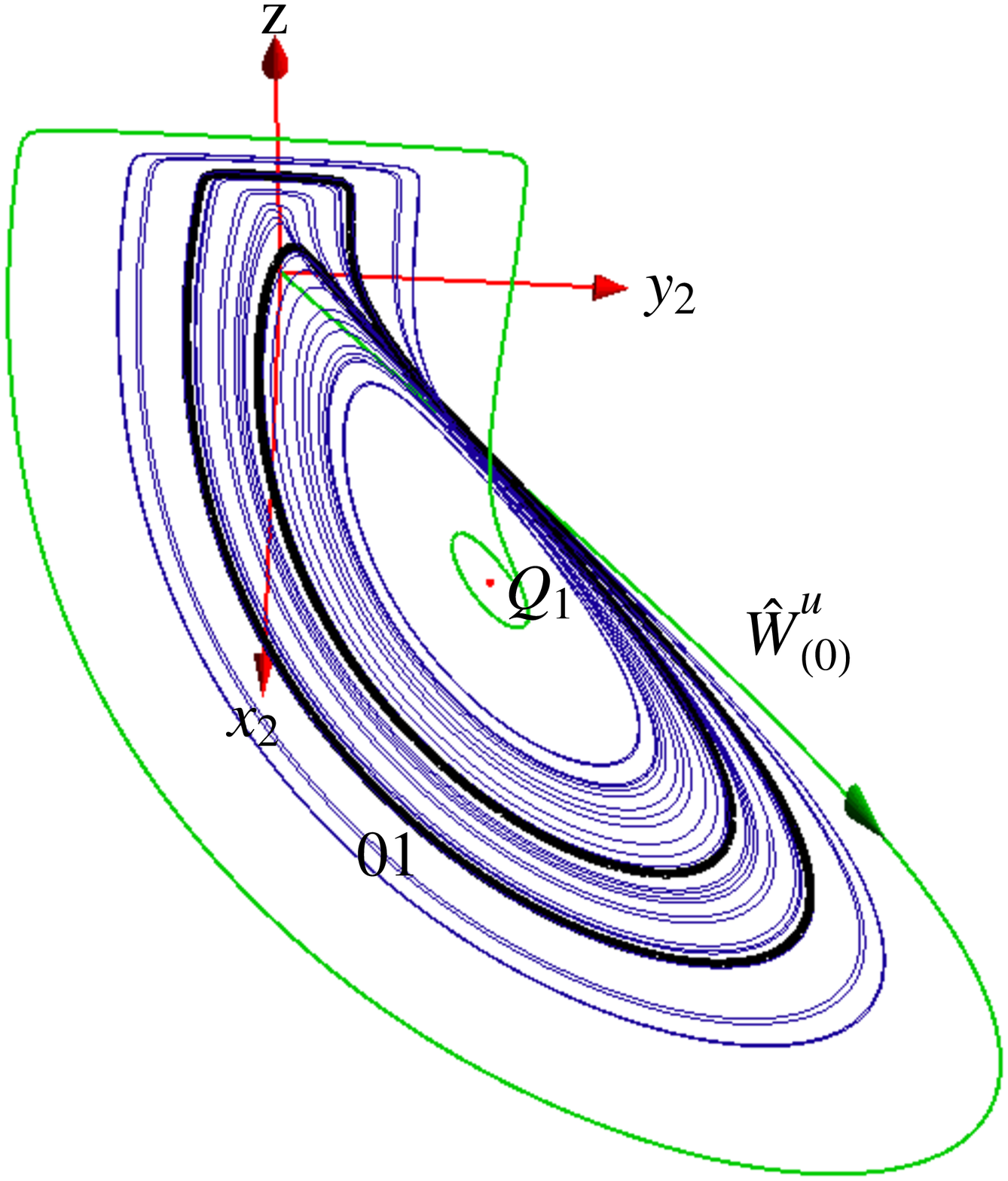}
 ~~~~(\textit{b})
\includegraphics[width=0.41\textwidth,clip=true]{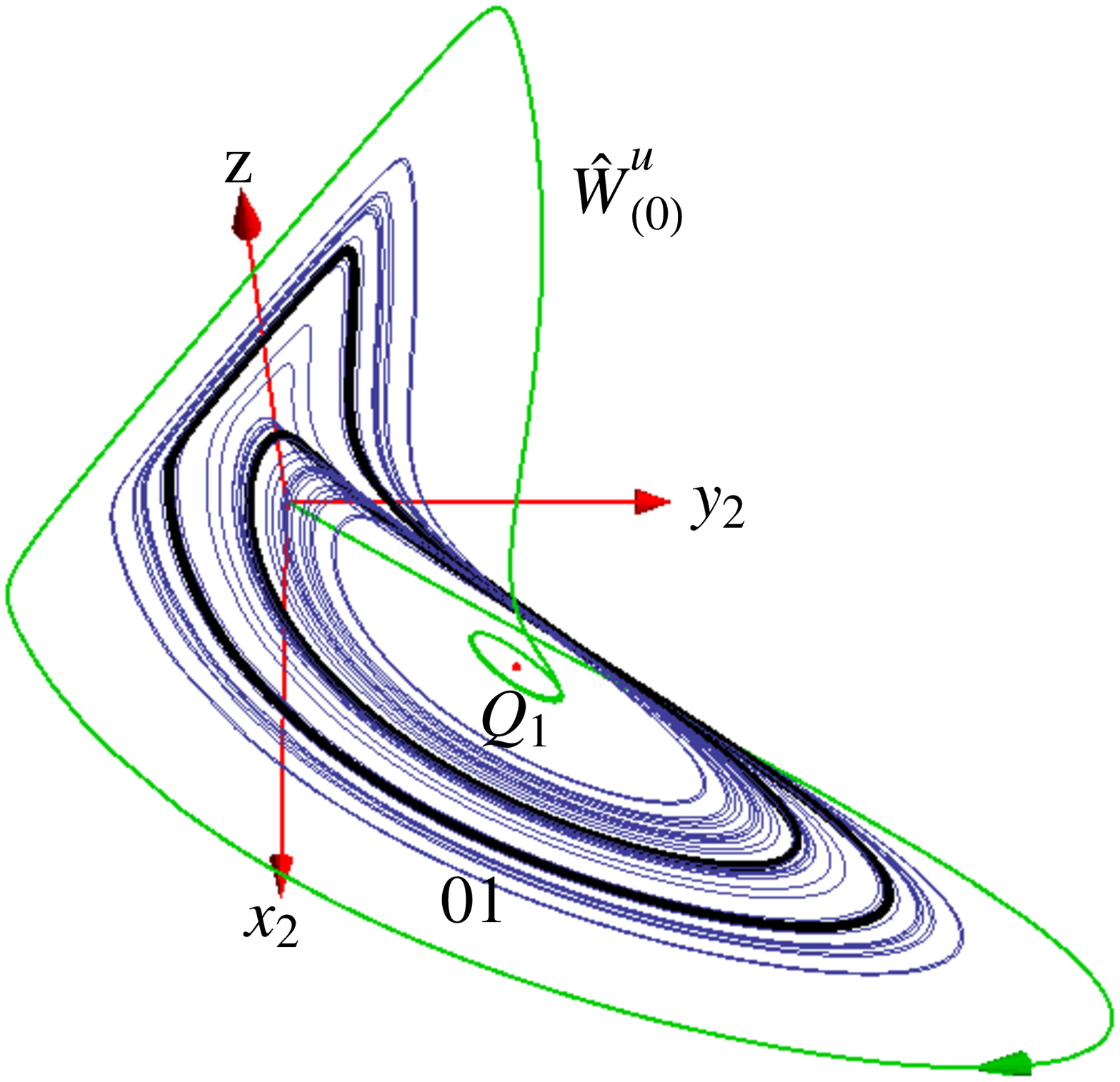}
\end{center}
\caption{
\Statesp\ portrait of \cLf\ in \reducedsp, projected on the
slice, taking as the slice fixing point $\slicep$
(a) the irreducible subspace \slice\ condition \refeq{cLeCoordSlice},
(b) the \reqv\ $\ssp_{\REQB{1}}$.
}
\label{fig:CLEmf}
\end{figure}
%%%%%%%%%%%%%%%%%%%%%%%%%%%%%%%%%%%%%%%%%%%%%%%%%%%%%%%%%%%%%%%%

As in \refsect{sec:CLeMovFr}, note that the in\-vari\-ants
are not well defined in the $x_1,\,x_2 \to 0$ limit.
Using $x=r_1\, e^{i\phi_1}\,,\, y=r_2\, e^{i\phi_2}$ we can write
\bea
	  \overline{x}_2 &=& r_1 \,,\qquad
	  \overline{y}_1 \,=\, r_2\sin(\phi_1-\phi_2)\continue
	  \overline{y}_2 &=& r_2\cos(\phi_1-\phi_2) \,,\qquad	
	  \overline{z} \,=\, z\,.
	  \label{eq:invLaserPolar}
\eea
For any given $y$ (therefore also for given $\phi_2$), the
limit of $\overline{y}$ for $x \rightarrow 0$ does not exist,
as the above expression depends on the direction in the
complex $x$-plane along which we approach zero.

From a different perspective, we may redefine the slice so
that $x_1=x_2=0$ is excluded, that is by
$\overline{x}_1=0,\;\overline{x}_2>0$. Then we may say that
group orbits of points in the $x_1=x_2=0$ subspace fail to
intersect the slice. In \reffig{fig:CLEmf}\,(a) this becomes
apparent by the trajectories in reduced space being stretched
as they come closer to $x_1=x_2=0$ subspace, where transverse
intersection would eventually fail.

It is instructive to rewrite the \cLe~\refeq{eq:CLe} in terms
of the invariant variables \refeq{eq:invLaser}. This is
achieved by using the chain rule \refeq{HilbChainRl} and
expressing the result in terms of variables
\refeq{eq:invLaser}. The moving frames symmetry-reduced \cLe\
are a 4-dimensional ODE system
\bea
	\dot{\overline{x}}_2 &=& -\sigma \,(\overline{x}_2 - \overline{y}_2)
\,,\quad
	\dot{\overline{y}}_1 \,=\, - \overline{y}_1 + \ImrCLor \overline{x}_2
  - \left(e + {\sigma\,\overline{y}_1}/{\overline{x}_2}\right) \overline{y}_2
\continue
	\dot{\overline{y}}_2 &=&  - \overline{y}_2 + (\RerCLor-z)\,\overline{x}_2
   + \left(e + {\sigma\,\overline{y}_1}/{\overline{x}_2}\right) \overline{y}_1
\,,\quad
	\dot{z}\; \,=\, -b\,z + \overline{x}_2 \overline{y}_2
\,.
\label{eq:rdcdCLeR}
\eea
Note the singularity as $\overline{x}_2=r_1\rightarrow 0$.

The projections in \reffig{fig:CLEmf} help us understand the
topology of the dynamics but also present large discontinuous jumps.
Note that the in\-vari\-ants \refeq{eq:invLaser} are related to
the in\-vari\-ant polynomials \refeq{eq:ipLaser} by division by
$\sqrt{x_1^2+x_2^2}$. This is the reason we get a clearer
visualization of the dynamics than with in\-vari\-ant
polynomials: All in\-vari\-ants scale as the original
coordinates. At the same time division by
$\sqrt{x_1^2+x_2^2}$ causes the jumps in the $\overline{y}$
components whenever the magnitude of $x$ comes close to zero.

So what does this imply for the ultimate goal of this paper,
``Continuous symmetry reduction and return maps for
higher-dimensional flows?'' In this example there is no
problem. It is obvious by inspection of \reffig{fig:CLEmf}
that one can chose a \Poincare\ section \emph{away} from the
singular set. Repeating the construction of
\refsect{s:Poincare} results in a return map very much like
the one of \reffig{fig:CLEip}\,(b), with the same admissible
orbits, so there is no need to plot further return maps for
various choices of \slice s.

\subsection{\label{s:mfReqb}\CLe: the general linear \slice}

The irreducible subspace \slice\ condition
\refeq{cLeCoordSlice} yields the in\-vari\-ant variables
\refeq{eq:invLaser} in explicit, analytic form. As explained
in \refsect{sec:CLeMovFr}, the \mframes\ also introduces
artificial singularities in \reducedsp, the location of which
depend on the choice of slice point, and one might suspect that
the singularities encountered by the strange attractor are
due to very special choice of the slice fixing condition. How
does a general choice of the slice fixing point affect the singular
set? In this setting it is no longer convenient to explicitly write
out transformations to in\-vari\-ant variables as we did in
\refsect{sec:mf}; we will implement the moving frame map
numerically, mapping computed trajectories to the \slice.
Even though the analytical computation of
in\-vari\-ants by the {\mframes} can be implemented by computer
algebra\rf{SiminosThesis} for system dimensions of the order
of $100$, it is both computationally prohibitive and utterly
unnecessary for symmetry reduction of very high dimensional
flows of order $100,000$ required for fully resolved $3$-D
fluid simulations\rf{GibsonPhD}.

\begin{figure}[ht]
\begin{center}
  (\textit{a})\includegraphics[width=0.35\textwidth,clip=true]{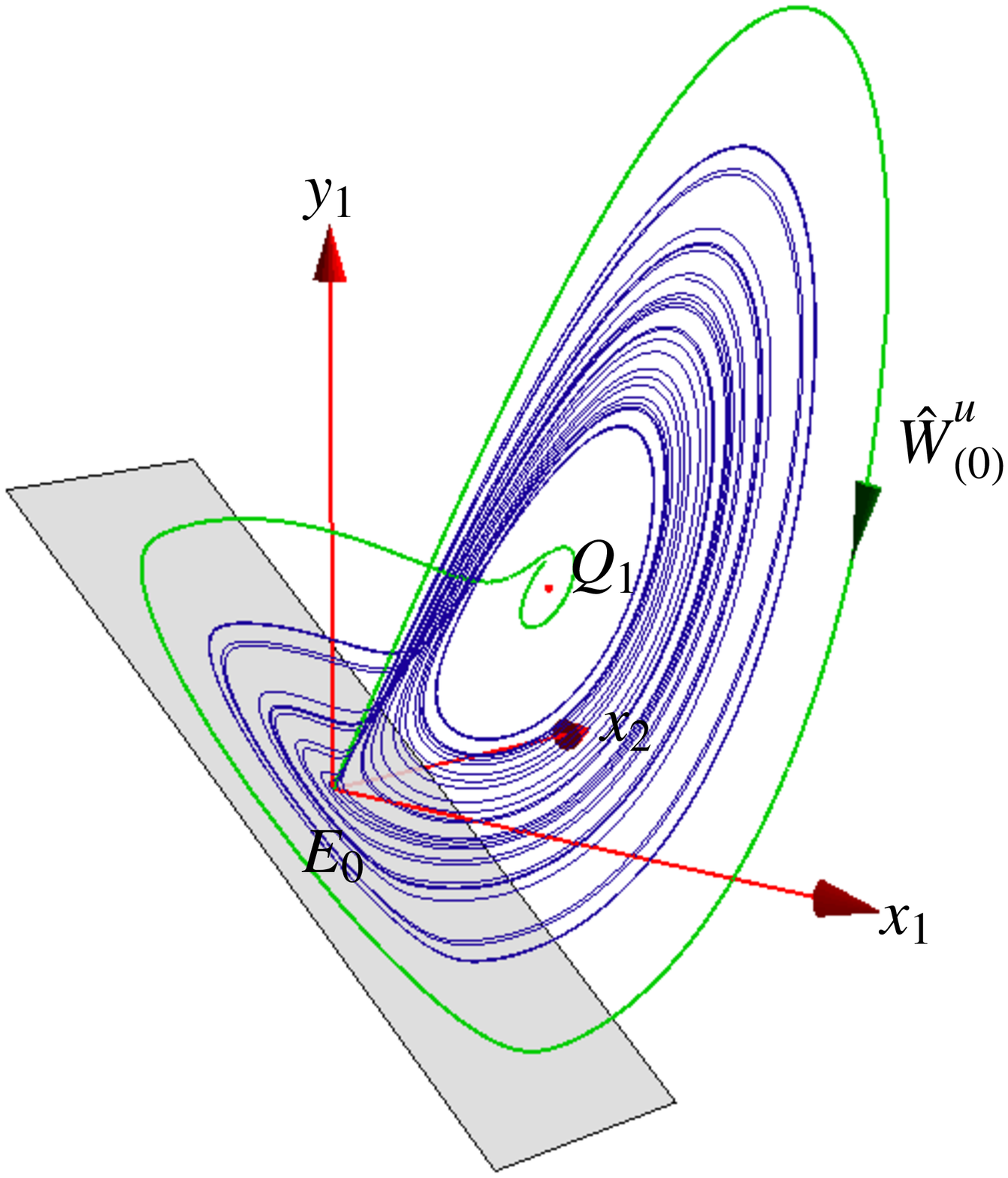}
~~~~(\textit{b})%
\includegraphics[width=0.35\textwidth,clip=true]{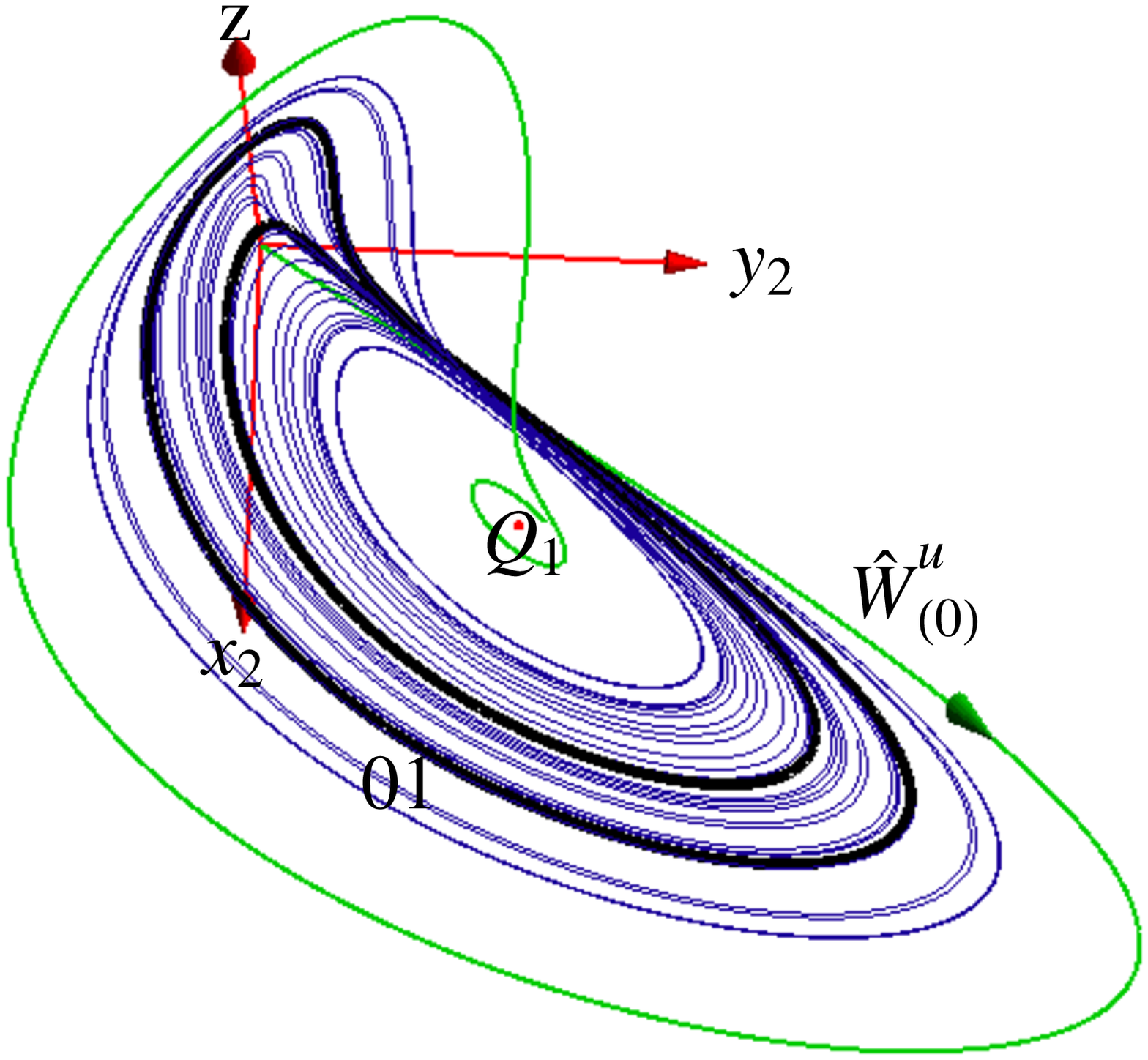}
\end{center}
\caption{
\Statesp\ portraits of \cLf\ in \reducedsp. We use a
moving frame map to a slice fixed by point
(a) $\slicep  = \ssp_{\REQB{1}}$, with
the gray plane indicating the \sset.
(b) $\slicep  = \ssp_{\REQB{1}}+(5,0,0,0,0)$. Compare with \reffig{fig:CLEmf}\,(b).
    }\label{fig:CLEmfsset}
\end{figure}

Since \REQB{1} organizes reduced space dynamics around it,
but also sets the scale of angular velocity of symmetry
induced rotations in the system, we find it natural to choose
slice-fixing point \refeq{CLEslicFixP} computed from
group orbit of the \reqv\ $\ssp_{\REQB{1}}$, given in \refeq{eq:Q1} for
the parameter values used here.
We can use
\refeq{cLeMF} to compute $\theta$ for any point $\ssp$, but
keeping up with the numerical approach of this section, we
use a Newton's method. As an initial guess we use the angle
from previous point along the trajectory, while for the first
point we choose among the two possible solutions by demanding
that the rotation brings the point to a minimum distance from
\slicep. Projections of \cLf\ to the slice defined in this
manner are shown in \reffig{fig:CLEmf}\,(b). This projection
of the strange attractor also
clearly exhibits the moving frame angle $\pi$-jumps.
\refFig{fig:CLEmfsset}(a) shows the projection of \sset\ in
$3$-dimensions $(x_1,x_2,y_1)$; the attractor
collides with the \sset\ head on.

The linear relation \refeq{cLeMFsingPerpTan} and
\reffig{fig:CLEmfsset} suggest that we can manipulate the
\sset\ so that the attractor avoids the singularity by
increasing the ratio $\slicepComp{y}{2}/\slicepComp{y}{1}$.
Choosing the slice fixing point as
$\slicep=\ssp_{\REQB{1}}+(5,0,0,0,0)$
we can `tilt' the \sset\ so that trajectories approach it in a
smoother manner, see \reffig{fig:CLEmfsset}\,(b).
%%%%%%%%%%%%%%%%%%%%%%%%%%%%%%%%%%%%%%%%%%%%%%%%%%%%%%%%%%%%%%%%
%%%%%%%%%%%%%%%%%%%%%%%%%%%%%%%%%%%%%%%%%%%%%%%%%%%%%%%%%%%%%%%%
%

For more pointers on how to pick good slices and combine them
with well-chosen \Poincare\ sections, the reader is referred to
\refref{SiminosThesis}.

\section{\label{sec:MovFrameODE}Differential formulation: the \mslices}

Instead of post-processing a full \statesp\ trajectory, we
can proceed as follows: Split up the integration into a
sequence of short time steps, each followed by a rotation of
the final point such that the next segment's initial point is
in the {\em \slice} fixed by a point $\slicep $.
In the infinitesimal steps limit
this leads to the \emph{\mslices}, a differential form of the
\mframes\ for which the trajectory never leaves the
\reducedsp.

Consider an $N$-dimensional Lie group $\Group$ acting on
$d$-dimensional space and which, at least locally near
$\slicep$, has $N$-dimensional orbits. For points that can be
mapped by a moving frame to \slice\ through $\slicep$ we can
write, using decomposition \refeq{EquiTraj}, the full
\statesp\ trajectory as $\ssp(\tau)=
\LieEl(\tau)\,\sspRed(\tau)$, where the
$(d\!-\!N)$-dim\-ens\-ion\-al \reducedsp\ trajectory
$\sspRed(\tau)$ is to be fixed by some condition, and
$\LieEl(\tau)$ is then the corresponding group action on the
$N$-dim\-ens\-ion\-al group manifold that
rotates $\sspRed$ into $\ssp$ at time $\tau$. The time
derivative is $\dot{\ssp}= \vel(\LieEl\sspRed) =
\dot{\LieEl}\sspRed + \LieEl\velRed$, with the \reducedsp\
velocity field given by $\velRed={d\sspRed}/{d\tau}$. Rewriting
this as $ \velRed = \LieEl^{-1} \vel(\LieEl \, \sspRed) -
\LieEl^{-1} \dot{\LieEl} \, \sspRed $ and using the
equivariance condition \refeq{eq:equivFinite} leads to
\[
\velRed = \vel - \LieEl^{-1} \dot{\LieEl} \, \sspRed
\,.
\]
The Lie group element \refeq{FiniteRot} and its time
derivative describe the group tangent flow
\[
\LieEl^{-1} \dot{\LieEl} =
\LieEl^{-1}\frac{d~}{d\tau} e^{\gSpace \cdot \Lg } =
\dot{\gSpace} \cdot \Lg
\,.
\]
This is the group tangent velocity $\LieEl^{-1} \dot{\LieEl}
\, \sspRed = \dot{\gSpace} \cdot \groupTan(\sspRed)$
evaluated at the point \sspRed, \ie, with ${\LieEl} = 1$,
see \reffig{fig:ReducTraj}\,(a).
 The flow in the $(d\!-\!N)$
directions transverse to the group flow is now obtained by
subtracting the flow along the group tangent direction,
\beq
\velRed(\sspRed) = \vel(\sspRed)
      - \dot{\gSpace}(\sspRed) \cdot \groupTan(\sspRed)
\,,\qquad
\velRed={d\sspRed}/{d\tau}
\,,
\ee{reducFlow}
for any factorization of the flow of form $\ssp(\tau)=
\LieEl(\tau)\, \sspRed(\tau)$. To integrate these equations
we first have to fix a particular flow factorization by
imposing conditions on $\sspRed(\tau)$, and then integrate
phases $\gSpace(\tau)$ on a given \reducedsp\ trajectory
$\sspRed(\tau)$.

Here we shall demand that the \reducedsp\ is confined to a
linear hyperplane \slice. Substituting \refeq{reducFlow} into the
time derivative of the fixed \slice\ condition
\refeq{PCsectQ1},
\[
\velRed(\sspRed)^T \sliceTan{a} =
\vel(\sspRed)^T \sliceTan{a} -
\dot{\gSpace}_a \cdot
\groupTan(\sspRed)^T  \sliceTan{a}
= 0
    \,,
\]
yields the equation for the group phases flow $\dot{\gSpace}$
for the \slice\ fixed by \slicep, together with the
\reducedsp\ $\pSRed$  flow $\velRed(\sspRed)$:
\bea
\dot{\gSpace}_a(\sspRed) &=& \frac{\vel(\sspRed)^T \sliceTan{a}}
                       {\groupTan(\sspRed)^T \cdot \sliceTan{} }
\label{MFdtheta}\\
\velRed(\sspRed) &=& \vel(\sspRed)
                    \,-\, \dot{\gSpace}(\sspRed)  \cdot \groupTan(\sspRed)
    \,,\qquad\quad \sspRed \in \pSRed
\,.
\label{EqMotMFrame}
\eea
Each group orbit $\pS_\ssp = \{  \LieEl \, \ssp \,|\, \LieEl
\in \Group \}$ is an equivalence class; the \mslices\ represents
the class by its single \slice\ intersection point $\sspRed$.
By construction $\velRed^T \sliceTan{a} = 0$, and  the motion
stays in the $(d\!-\!N)$-dim\-ens\-ion\-al \slice. We have
thus replaced the original dynamical system $\{\pS,f\}$ by a
reduced system $\{\pSRed,\bar{f}\}$.

These equations are easily integrated (provided some care is
taken about how the trajectories cross the singular set), and
given the same slice-fixing conditions, the integrations reproduce
the plots obtained by the \mframes, such as \reffig{fig:CLEmf}.

For example, consider the \cLe\ \slice\ condition of
\refsect{s:cleCoordSlice}: $x_1=0,\;x_2>0$. The \reducedsp\
equations are given by
\beq
\velRed(\sspRed) =
   \vel(\sspRed) - \frac{\vel_1}{\sspRed_2} \groupTan(\sspRed)
\,.
\ee{ExerMotionMovFrame}
Substitution of the \cLe\ equations recovers \refeq{eq:rdcdCLeR},
obtained by the \mframes. The integration of
\refeq{eq:rdcdCLeR} recovers the strange attractor in
\reffig{fig:CLEmf}\,(a), obtained by simulation in the full
\statesp, followed by moving frame rotation into the \slice.

In pattern recognition and `template fitting' settings,
\refeq{MFdtheta} is called the {\em reconstruction equation}.
We have already encountered it in our polar coordinates
exercise \refeq{eq:PolarCLeAngles}. Integrated together, the
\reducedsp\ trajectory \refeq{EqMotMFrame}, the integrated
phase \refeq{MFdtheta}, and $\LieEl(\tau)=\exp[\gSpace(\tau)
\cdot \Lg]$ reconstruct the full \statesp\ trajectory
$\ssp(\tau)= \LieEl(\tau)\,\sspRed(\tau)$ from the
\reducedsp\  trajectory $\sspRed(\tau)$, so no information
about the flow is lost in the process of symmetry reduction.

The denominator in \refeq{MFdtheta} vanishes and the phase
velocity $\dot{\gSpace}(\sspRed)$ diverges whenever the
direction of group action on the \reducedsp\ point is
perpendicular to the direction of group action on the \slice\
point $\slicep$. Therefore the \mslices\ has the same \sset\ as
its post-processing variant, the \mframes: the intersection of
the slice with the set of points with group tangent
perpendicular to $\sliceTan{}$.

\emph{A historical note.}
The basic idea of the \mslices\ is intuitive and
frequently reinvented, often under a different name; for example,
it is stated without attribution as problem 1. of Sect.
6.2 of Arnol'd {\em Ordinary Differential
Equations}\rf{arnold92}. The factorization
\refeq{EquiTraj} is stated on p.~31 of Anosov and
Arnol'd\rf{AnAr88}, who note, without further elaboration,
that in the vicinity of a point that is not fixed by the
group one can reduce the order of a system of differential
equations by the dimension of the group.
Fiedler, in the influential 1995 talk
at the Newton Institute, and Fiedler, Sandstede, Wulff,
Turaev and  Scheel\rf{FiSaScWu96,SaScWu97,SaScWu99a,FiTu98}
treat Euclidean symmetry bifurcations in the context of
spiral wave formation. The central idea is to utilize the
semidirect product structure of the Euclidean group $E(2)$ to
transform the flow into a `skew product' form, with a part
orthogonal to the group orbit, and the other part within it,
as in \refeq{EqMotMFrame}. They refer to a linear slice
\pSRed\ near a \reqv\ as a {\em Palais slice}, with Palais
coordinates. As the choice of the slice is arbitrary, these
coordinates are not unique. According to these authors, the
skew product flow was first written down by
Mielke\rf{Mielke91}, in the context of buckling in
elasticity theory. However, this decomposition is no doubt
much older. For example, it was  used by
Krupa\rf{Krupa90,ChossLaut00} in his local slice study of
bifurcations of \reqva. Biktashev, Holden, and
Nikolaev\rf{BiHoNi96} cite Anosov and Arnol'd\rf{AnAr88}  for
the `well-known' factorization \refeq{EquiTraj} and write
down the slice flow equations \refeq{EqMotMFrame}.
Haller and Mezi\'c\rf{HaMe98} reduce symmetries of
three-dimensional volume preserving flows and reinvent
\mframes, under the name `orbit projection map.' There is
extensive literature on reduction of symplectic manifolds
with symmetry; the Marsden and Weinstein 1974 article\rf{MaWe74}
is an important early reference. Then there are studies of
the reduced phase spaces for vortices moving on a sphere such
as \refref{Kirwan88}, and many, many others.

Neither Fiedler \etal\rf{FiSaScWu96} nor Biktashev
\etal\rf{BiHoNi96} implemented their methods numerically.
That was done by Rowley and Marsden for the
Kuramoto-Sivashinsky\rf{rowley_reconstruction_2000} and the
Burgers\rf{rowley_reduction_2003} equations, and Beyn and
Th\"ummler\rf{BeTh04,Thum05} for a number of
reaction-diffusion systems, described by parabolic partial
differential equations on unbounded domains. We recommend the
Barkley paper\rf{Barkley94} for a clear explanation of how
the Euclidean symmetry leads to spirals, and the Beyn and
Th\"ummler paper\rf{BeTh04} for inspirational concrete
examples of how freezing/\-slicing simplifies the
dynamics of rotational, traveling and spiraling \reqva.

Beyn and Th\"ummler write the solution as a composition of
the action of a time dependent group element $\LieEl(\tau)$ with
a `frozen,' in-\slice\ solution $\hat{u}(\tau)$
\refeq{EquiTraj}. In their nomenclature, making a \reqv\ stationary
by going to a co-moving frame is `freezing' the
traveling wave, and the imposition of the phase
condition (\ie, \slice\ condition \refeq{PCsectQ}) is the
`freezing ansatz.' They find it more convenient to make use
of the equivariance by extending the \statesp\ rather than
reducing it, by adding an additional parameter and a phase
condition.
The freezing ansatz\rf{BeTh04} is identical
to the Rowley and Marsden\rf{rowley_reduction_2003} and our
slicing, except that freezing is formulated as an
additional constraint, just as when we compute periodic
orbits of ODEs we add Poincar\'e section as an additional
constraint, \ie, increase the dimensionality of the problem
by 1 for every continuous symmetry.

Our derivation of \mslices\ follows most closely
Rowley and Marsden\rf{rowley_reduction_2003} who in the
pattern recognition setting refer to the \slice\ point as a
{\em template}, and call \refeq{MFdtheta} the {\em reconstruction
equation}\rf{MarsdRat94}. They also describe the `method
of connections' (called `orthogonality of time and group
orbit at successive times' in \refref{BeTh04}), for which the
reconstruction equation \refeq{MFdtheta} denominator is
$\groupTan(\sspRed)^T \cdot \groupTan(\sspRed)$ and thus
nonvanishing as long as the action of the group is regular.
This avoids the spurious \slice\ singularities, but it is not
clear what the method of connections buys us otherwise. It
does not reduce the dimensionality of the \statesp, and it
accrues geometric phases which prevent \rpo s from closing
into \po s.
Geometric phase in laser equations, including \cLe,
has been studied in \refref{ToDe94,ToDe94a,NiHa91,NiHa92,NiHa92a}.

One would think that with all the literature on
desymetrization the case is shut and closed, but not so.
Applied mathematicians are inordinately fond of bifurcations,
and almost all of the previous work focuses on \eqva, \reqva\
(traveling waves), and their bifurcations, and for these
problems a single \slice\ works well. Only when one tries to
describe the totality of chaotic orbits does the non-global
nature of slices become a serious nuisance.

\section{\label{sec:concl} Discussion and conclusions}

We have presented two approaches to continuous symmetry
reduction of higher-dimensional flows and illustrated them
with reductions of the complex Lorenz system, a 5-dimensional
dissipative flow with rotational symmetry.
In either approach numerical computations can be performed in
the original, full state-space representation, and then the
solutions can be projected onto the symmetry-reduced state
space.

In the {\em Hilbert polynomial basis} approach, one transforms
the equi\-vari\-ant \statesp\ coordinates into in\-vari\-ant coordinates by
a nonlinear coordinate transformation
\[
\{x_1,x_2,\cdots,x_d\} \to \{u_1,u_2,\cdots,u_m\}
\,,
\]
and studies the in\-vari\-ant image of dynamics rewritten in
terms of in\-vari\-ant coordinates. These in\-vari\-ant polynomial
bases can be algorithmically determined for both Hamiltonian
and dissipative systems. Our goal is to reduce symmetries of
fully-resolved simulations of PDEs, with state space
dimensions of the order of a few tens to few hundreds (for
\KS\ flow), and well into the tens or hundreds of thousands (for pipe
and \pCf s). Unfortunately, the computational cost of
polynomial basis algorithms is at present prohibitive for
state space dimensions larger than ten, so the in\-vari\-ant
polynomial basis approach is not a feasible option. We have
discussed it here solely for illustrative purposes.

In the \emph{\mframes} (or its continuous time, differential
version, the \emph{\mslices}), one fixes a local slice
$(\sspRed - \slicep )^T \groupTan' =0$, a hyperplane normal
to the group tangent \sliceTan{} that cuts across group
orbits in the neighborhood of the slice-fixing point
$\slicep$. The \statesp\ is sliced locally in such a way
that each group orbit of symmetry-equivalent points is
represented by a single point, with the symmetry-reduced
dynamics in the \reducedsp\ $\pS/\Group$ given by
\refeq{EqMotMFrame}:
\[
\velRed = \vel - \dot{\gSpace}  \cdot \groupTan
    \,,\qquad
\dot{\gSpace} = (\vel \cdot \groupTan')/(\groupTan \cdot \groupTan')
\,.
\]
The {\mframes}
turns out to be an efficient method for reducing the
flow to a symmetry-in\-vari\-ant \reducedsp, suited to reduction
of {even very high-dimensional}
dissipative flows to local return maps:
one runs the dynamics in the full \statesp\ and
post-processes the trajectory by the \mframes.
Importantly, from a numerical point of  view,
there is no need to
actually recast the dynamics in the new coordinates or
write new code. Either
approach can be used as a visualization tool, with all
computations carried out in the original coordinates, and
then, when done, projecting the solutions onto the symmetry
\reducedsp\ by post-processing the data.
In contrast to co-moving frames local to each traveling
solution, {restricting the dynamics}
to a slice renders all \reqva\
stationary in the same set of coordinates.

An inconvenience inherent in the linear slices formulation is that they are
local, and the reduced flow encounters singularities in
subsets of the \reducedsp, with the reduced trajectory exhibiting
large, slice-induced jumps.
This \sset\ is introduced by and
depends on the \slice-fixing condition. We have shown,
in the $5$-dimensional \cLe\ example, that the location of
the \sset\ can be manipulated by judicious choice of the slice
fixing point, and geometrical information about the dynamics can
be extracted by constructing a return map through a
\Poincare\ section that does not intersect the singular set.
The trick is to
construct a good set of symmetry in\-vari\-ant Poincar\'e
sections, and that is a dark art for systems of
dimension higher than three, with or without a
symmetry.
In higher-dimensional flows, with more involved symmetry
group actions and larger sets of stationary solutions, where a
single slice and \Poincare\ section will not suffice, we can
still expect to cover the \reducedsp\ with multiple slices,
obtaining a set of discrete maps involving multiple
\Poincare\ sections. As to high-dimensional applications,
it was shown in \refref{SiminosThesis} that the coexistence
of four equilibria, two \reqva\ and a nested \fixedsp\
structure in an effectively $8$-dimensional \KS\
system\rf{SCD07} complicates matters considerably. This
application of symmetry reduction to a spatially extended,
PDE system is the
subject of a forthcoming publication\rf{SCD09b}.

\section*{Acknowledgements}
We sought in vain Lou Howard's sage counsel on how
to desymmetrize, but none was forthcoming - hence this article.
We are, however, grateful to
D.~Barkley,
W.-J.~Beyn,
R.~Gilmore,
J.~Halcrow,
K.A.~Mitchell,
C.W.~Rowley,
R.~Wilczak,
and in particular R.L.~Davidchack for many spirited exchanges,
and J.F.~Gibson for a critical reading of the manuscript.
P.C. thanks the
James Franck Institute, U. of Chicago,
for hospitality, and Argonne National Laboratory and
G.~Robinson Jr. for partial support.
E.S. was supported by NSF grant DMS-0807574 and G.~Robinson,~Jr.

\bibliographystyle{elsarticle-num}

\end{document}